\documentclass[aps,prx,twocolumn,floatfix,10pt,amssymb,amsfont,amsmath,superscriptaddress,float,nofootinbib,longbibliography]{revtex4-2}

\usepackage{preamb}

\begin{document}
\pgfdeclarelayer{back}
\pgfdeclarelayer{front}
\pgfsetlayers{back,main,front}

\def\fixboundingbox{
    \path [use as bounding box] (current bounding box.south west) rectangle (current bounding box.north east);
}

\tikzset{->-/.style={decoration={
                markings,
                mark=at position #1 with {\arrow{Triangle[width=3.5pt, length=3.5pt]}}},postaction={decorate}}}

\tikzset{-<-/.style={decoration={
                markings,
                mark=at position #1 with {\arrowreversed{Triangle[width=3.5pt, length=3.5pt]}}},postaction={decorate}}}

\tikzset{
    every picture/.append style={
        line join=round,
        line width = 0.7pt,
        line cap=round
    }
}

\tikzset{mpoint/.style={ transform shape, sloped, 
minimum width=20pt, minimum height=4pt, inner sep=0pt, ultra thin, font=\tiny}}

\tikzset{
    clip even odd rule/.code={\pgfseteorule},
    invclip/.style={clip,insert path=[clip even odd rule]{
    [reset cm](-\maxdimen,-\maxdimen)rectangle(\maxdimen,\maxdimen)
    }}}

\newcommand\clipIntersection[1]{
    (#1.north west) -- (#1.north east) -- (#1.south east) -- (#1.south west) -- (#1.north west)
}

\tikzstyle{dot}=[dash pattern= on 0.6pt off 2.1pt]
\tikzstyle{mor}=[circle, minimum size=5pt, inner sep=1pt, fill=gray!20]
\tikzstyle{mps}=[circle, inner sep=2.5pt, fill=gray!20, draw]
\tikzstyle{mat}=[circle, inner sep=2pt, fill=gray!20, draw]
\tikzstyle{cg}=[circle, inner sep=4pt, fill=gray!20, draw]
\tikzstyle{mpo}=[rectangle, rounded corners=2pt, inner sep=3.5pt, fill=gray!20, draw]
\tikzstyle{obj1}=[draw=black]
\tikzstyle{obj2}=[draw=LimeGreen]
\tikzstyle{obj3}=[draw=BrickRed]


\newcommand\clipAroundNode[2]{
    \begin{pgfscope}
        \begin{scope}[overlay]
            \path[invclip] \clipIntersection{#1};#2
        \end{scope}
    \end{pgfscope}
}

\newcommand\clipAroundTwoNodes[3]{
    \begin{pgfscope}
        \begin{scope}[overlay]
            \path[invclip] \clipIntersection{#1} \clipIntersection{#2};#3
        \end{scope}
    \end{pgfscope}
}

\newcommand\clipAroundThreeNodes[4]{
    \begin{pgfscope}
        \begin{scope}[overlay]
            \path[invclip] \clipIntersection{#1} \clipIntersection{#2} \clipIntersection{#3};#4
        \end{scope}
    \end{pgfscope}
}

\newcommand\clipAroundFourNodes[5]{
    \begin{pgfscope}
        \begin{scope}[overlay]
            \path[invclip] \clipIntersection{#1} \clipIntersection{#2} \clipIntersection{#3} \clipIntersection{#4};#5
        \end{scope}
    \end{pgfscope}
}

\newcommand{\middleCoo}[6]{
    \path (#1) --coordinate[pos=0.5](#3) (#2);
    \path (#1) --coordinate[pos={0.5-#6}](#4) (#2);
    \path (#1) --coordinate[pos={0.5+#6}](#5) (#2);
}

\newcommand{\barycentre}[8]{
    \middleCoo{#1}{#2}{A}{}{}{#8};
    \middleCoo{#2}{#3}{B}{}{}{#8};
    \middleCoo{#1}{#3}{C}{}{}{#8};
    \path[name path=PA#3] (A) -- (#3);
    \path[name path=PB#1] (B) -- (#1);
    \path[name path=PC#2] (C) -- (#2);
    \path [name intersections={of=PA#3 and PB#1,by={#4}}];
    \path (#1) --coordinate[pos=0.84](#5) (#4);
    \path (#2) --coordinate[pos=0.84](#6) (#4);
    \path (#3) --coordinate[pos=0.84](#7) (#4);
}

\newcommand{\barycentreSq}[9]{
    \middleCoo{#1}{#2}{A1}{A2}{A3}{\sp};
    \middleCoo{#3}{#4}{B1}{B2}{B3}{\sp};
    \middleCoo{#1}{#3}{C1}{C2}{C3}{\sp};
    \middleCoo{#2}{#4}{D1}{D2}{D3}{\sp};
    \middleCoo{A1}{B1}{#5}{}{}{\sp};
    \path[name path=PA2B2] (A2) -- (B2);
    \path[name path=PA3B3] (A3) -- (B3);
    \path[name path=PC2D2] (C2) -- (D2);
    \path[name path=PC3D3] (C3) -- (D3);       
    \path [name intersections={of=PA2B2 and PC2D2,by={#6}}];
    \path [name intersections={of=PA2B2 and PC3D3,by={#8}}];
    \path [name intersections={of=PA3B3 and PC2D2,by={#7}}];
    \path [name intersections={of=PA3B3 and PC3D3,by={#9}}];
}

\newcommand{\intersec}[5]{
    \path[name path=P12] (#1) -- (#2);
    \path[name path=P34] (#3) -- (#4);
    \path [name intersections={of=P12 and P34,by={#5}}];
}

\newcommand{\mor}[2]{
    \node[mor, circle, inner sep=3.3pt] at (#1) {};
    \node at (#1) {${\sss #2}$};
}

\newcommand{\centerarc}[5]{
    \draw[#1] ($(#2)+({#5*cos(#3)},{#5*sin(#3)})$) arc (#3:#4:#5);
}


\newcommand{\MPS}[1]{
    \begin{tikzpicture}[scale=1,baseline={([yshift=-.55ex]current bounding box.center)}]
        \def\l{.6};
        \def\sh{1.23};
        \draw[obj2, -<-=.47] (\sh*4*\l,0) -- (\sh*4*\l+\l,0);
        \foreach \x in {4,3,2,1,0}{
            \draw[obj2, ->-=.75] (\sh*\x*\l,0) -- (\sh*\x*\l-\l,0);
            \draw[obj1, ->-=.75] (\sh*\x*\l,0) -- (\sh*\x*\l,\l);
            \node[mps] at (\sh*\x*\l,0) {};
            \node[yshift=-8pt] at (\sh*\x*\l,0) {${\sss #1}$};
        }
    \end{tikzpicture}
}

\newcommand{\MPSW}[1]{
    \begin{tikzpicture}[scale=1,baseline={([yshift=-.85ex]current bounding box.center)}]
        \def\l{.6};
        \def\sh{1.23};
        \draw[obj2, -<-=.47] (\sh*4*\l,0) -- (\sh*4*\l+\l,0);
        \foreach \x in {4,3,2,1,0}{
            \draw[obj2, ->-=.75] (\sh*\x*\l,0) -- (\sh*\x*\l-\l,0);
            \draw[obj1, ->-=.75] (\sh*\x*\l,0) -- (\sh*\x*\l,\l);
            \node[mps] at (\sh*\x*\l,0) {};
        }
        \ifthenelse{#1=1}{
            \foreach \x in {4,3,1,0}{
                \node[yshift=-8pt] at (\sh*\x*\l,0) {${\sss \rho_P}$};
            }
            \node[yshift=-8pt] at (\sh*2*\l,0) {${\sss \rho_S}$};
            \node[mat] at (2.5*\sh*\l,0) {};
            \node[mat] at (1.5*\sh*\l,0) {};
            \node[yshift=7pt] at (2.5*\sh*\l,0) {${\sss \pi}$};
            \node[yshift=7pt] at (1.5*\sh*\l,0) {${\sss \iota}$};
        }{}
        \ifthenelse{#1=2}{
            \node[yshift=-8pt] at (\sh*4*\l,0) {${\sss \rho_P}$};
            \node[yshift=-8pt] at (0,0) {${\sss \rho_P}$};
            \node[yshift=-8pt] at (\sh*1*\l,0) {${\sss \rho_S}$};
            \node[yshift=-8pt] at (\sh*2*\l,0) {${\sss \rho_S}$};
            \node[yshift=-8pt] at (\sh*3*\l,0) {${\sss \rho_S}$};
            \node[mat] at (3.5*\sh*\l,0) {};
            \node[mat] at (.5*\sh*\l,0) {};
            \node[yshift=7pt] at (3.5*\sh*\l,0) {${\sss \pi}$};
            \node[yshift=7pt] at (.5*\sh*\l,0) {${\sss \iota}$};
        }{}
    \end{tikzpicture}
}

\newcommand{\MPST}[6]{
    \begin{tikzpicture}[scale=1,baseline=-.55ex]
        \def\l{.6};
        \def\sh{1.23};
        \draw[obj2, -<-=.5] (0,0) --node[pos=1.25]{${\sss #5}$} (\l,0);
        \draw[obj2, ->-=.76] (0,0) --node[pos=1.25]{${\sss #6}$} (-\l,0);
        \draw[obj1, ->-=.75] (0,0) --node[pos=1.25]{${\sss #2}$} (0,\l);
        \node[mps] at (0,0) {};
        \node[yshift=-8pt] at (0,0) {${\sss #1}$};
        \node[yshift=-6pt] at (-\sh*\l/2,0) {${\sss #3}$};
        \node[yshift=-6pt] at (\sh*\l/2,0) {${\sss #4}$};
    \end{tikzpicture}
}

\newcommand{\actionMPO}[6]{
    \begin{tikzpicture}[scale=1,baseline={([yshift=-.55ex]current bounding box.center)}]
        \def\l{.6};
        \def\sh{1.23};
        \ifthenelse{#1=1}{
            \draw[obj2, -<-=.47] (0,0) -- (\l,0);
            \draw[obj2, ->-=.75] (0,0) -- (-\l,0);
            \draw[obj3, -<-=.47] (0,\sh*\l) -- (\l,\sh*\l);
            \draw[obj3, ->-=.75] (0,\sh*\l) -- (-\l,\sh*\l);
            \draw[obj1, ->-=.75] (0,0) -- (0,\l);
            \draw[obj1, ->-=.75] (0,\sh*\l) -- (0,\sh*\l+\l);
            \node[mps] at (0,0) {};
            \node[mpo] at (0,\sh*\l) {};
            \node[yshift=-8pt] at (0,0) {${\sss #2}$};
            \node[xshift=8pt, yshift=6pt] at (0,\sh*\l) {${\sss #3}$};
            \node[yshift=-6pt] at (-\sh*\l/2,0) {${\sss #6}$};
            \node[yshift=-6pt] at (\sh*\l/2,0) {${\sss #6}$};
        }{}
        \ifthenelse{#1=2}{
            \draw[obj2, -<-=.47] (0,0) -- (\l,0);
            \draw[obj2, -<-=.47] (\sh*\l,0) -- (\sh*\l+\l,0);
            \draw[obj2, ->-=.75] (0,0) -- (-\l,0);
            \draw[obj2, ->-=.75] (-\sh*\l,0) -- (-\sh*\l-\l,0);
            \draw[obj3, -<-=.72, rounded corners=2pt] (\sh*\l,0) -- (\sh*\l,\sh*\l) -- (\sh*\l+\l,\sh*\l);
            \draw[obj3, ->-=.85, rounded corners=2pt] (-\sh*\l,0) -- (-\sh*\l,\sh*\l) -- (-\sh*\l-\l,\sh*\l);
            \draw[obj1, ->-=.7] (0,0) -- (0,\sh*\l);
            \node[mps] at (0,0) {};
            \node[mps] at (\sh*\l,0) {};
            \node[mps] at (-\sh*\l,0) {};
            \node[yshift=-8pt] at (0,0) {${\sss #2}$};
            \node[xshift=7pt, yshift=7pt] at (\sh*\l,0) {${\sss #4}$};
            \node[xshift=-6pt, yshift=7pt] at (-\sh*\l,0) {${\sss #5}$};
            \node[yshift=-6pt] at (-3*\sh*\l/2,0) {${\sss #6}$};
            \node[yshift=-6pt] at (3*\sh*\l/2,0) {${\sss #6}$};
        }{}
    \end{tikzpicture}
}

\newcommand{\MPO}[1]{
    \begin{tikzpicture}[scale=1,baseline={([yshift=-.55ex]current bounding box.center)}]
        \def\l{.6};
        \def\sh{1.23};
        \draw[obj3, -<-=.47] (\sh*4*\l,0) -- (\sh*4*\l+\l,0);
        \foreach \x in {4,3,2,1,0}{
            \draw[obj3, ->-=.75] (\sh*\x*\l,0) -- (\sh*\x*\l-\l,0);
            \draw[obj1, ->-=.75] (\sh*\x*\l,0) -- (\sh*\x*\l,\l);
            \draw[obj1, -<-=.47] (\sh*\x*\l,0) -- (\sh*\x*\l,-\l);
            \node[mpo] at (\sh*\x*\l,0) {};
            \node[xshift=8pt, yshift=6pt] at (\sh*\x*\l,0) {${\sss #1}$};
        }
    \end{tikzpicture}
}

\newcommand{\MPOT}[5]{
    \begin{tikzpicture}[scale=1,baseline={([yshift=-.55ex]current bounding box.center)}]
        \def\l{.6};
        \draw[obj3, -<-=.47] (0,0) --node[pos=1.25]{${\sss #3}$} (\l,0);
        \draw[obj3, ->-=.75] (0,0) --node[pos=1.25]{${\sss #4}$} (-\l,0);
        \draw[obj1, ->-=.75] (0,0) --node[pos=1.25]{${\sss #2}$} (0,\l);
        \draw[obj1, -<-=.47] (0,0) --node[pos=1.25]{${\sss #5}$} (0,-\l);
        \node[mpo] at (0,0) {};
        \node[xshift=8pt, yshift=6pt] at (0,0) {${\sss #1}$};
    \end{tikzpicture}
}

\newcommand{\MPOTSpe}[5]{
    \begin{tikzpicture}[scale=1,baseline={([yshift=-.25ex]current bounding box.center)}]
        \def\l{.6};
        \draw[obj3, -<-=.47] (0,0) --node[pos=1.25]{${\sss #3}$} (\l,0);
        \draw[obj3, ->-=.75] (0,0) --node[pos=1.25]{${\sss #4}$} (-\l,0);
        \draw[obj1, ->-=.75] (0,0) --node[pos=1.25]{${\sss #2}$} (0,\l);
        \draw[obj1, -<-=.47] (0,0) --node[pos=1.25]{${\sss #5}$} (0,-\l);
        \node[mpo] at (0,0) {};
        \node[xshift=8pt, yshift=6pt] at (0,0) {${\sss #1}$};
        \node[yshift=-5pt] at (0,-\l) {${\sss \sigma^z}$};
    \end{tikzpicture}
}

\newcommand{\fusionT}[4]{
    \begin{tikzpicture}[scale=1,baseline={([yshift=-.55ex]current bounding box.center)}]
        \def\l{.6};
        \def\sh{1.23};
        \draw[obj3, -<-=.62, rounded corners=2pt] (0,0) -- (0,\sh*\l/2) -- node[pos=1.25]{${\sss #2}$} (\l,\sh*\l/2);
        \draw[obj3, -<-=.62, rounded corners=2pt] (0,0) -- (0,-\sh*\l/2) -- node[pos=1.25]{${\sss #3}$} (\l,-\sh*\l/2);
        \draw[obj3, ->-=.75] (0,0) --node[pos=1.25]{${\sss #4}$} (-\l,0);
        \node[mps] at (0,0) {};
        \node[xshift=11pt] at (0,0) {${\sss #1}$};
    \end{tikzpicture}
}

\newcommand{\actionT}[4]{
    \begin{tikzpicture}[scale=1,baseline={([yshift=-2.4ex]current bounding box.center)}]
        \def\l{.6};
        \def\sh{1.23};
        \draw[obj3, -<-=.72, rounded corners=2pt] (0,0) -- (0,\sh*\l) -- node[pos=1.25]{${\sss #2}$} (\l,\sh*\l);
        \draw[obj2, ->-=.75] (0,0) -- (-\l,0);
        \draw[obj2, -<-=.47] (0,0) -- (\l,0);
        \node[mps] at (0,0) {};
        \node[xshift=7pt, yshift=7pt] at (0,0) {${\sss #1}$};
        \node[yshift=-6pt] at (-\sh*\l/2,0) {${\sss #3}$};
        \node[yshift=-6pt] at (\sh*\l/2,0) {${\sss #4}$};
    \end{tikzpicture}
}

\newcommand{\actionTGen}[6]{
    \begin{tikzpicture}[scale=1,baseline={([yshift=-2.4ex]current bounding box.center)}]
        \def\l{.6};
        \def\sh{1.23};
        \draw[obj3, -<-=.72, rounded corners=2pt] (0,0) -- (0,\sh*\l) -- node[pos=1.25]{${\sss #2}$} (\l,\sh*\l);
        \draw[obj2, ->-=.75] (0,0) --node[pos=1.25]{${\sss #5}$} (-\l,0);
        \draw[obj2, -<-=.47] (0,0) --node[pos=1.25]{${\sss #6}$} (\l,0);
        \node[mps] at (0,0) {};
        \node[xshift=12pt, yshift=8pt] at (0,0) {${\sss #1}$};
        \node[yshift=-6pt] at (-\sh*\l/2,0) {${\sss #3}$};
        \node[yshift=-6pt] at (\sh*\l/2,0) {${\sss #4}$};
    \end{tikzpicture}
}

\newcommand{\orthoFusion}[0]{
    \begin{tikzpicture}[scale=1,baseline={([yshift=-.55ex]current bounding box.center)}]
        \def\l{.6};
        \def\sh{1.23};
        \def\r{2.5};
        \draw[obj3, -<-=.46, rounded corners=2pt] (0,0) -- (0,\sh*\l/2) -- (\r*\sh*\l,\sh*\l/2) -- (\r*\sh*\l,0);
        \draw[obj3, -<-=.46, rounded corners=2pt] (0,0) -- (0,-\sh*\l/2) -- (\r*\sh*\l,-\sh*\l/2) -- (\r*\sh*\l,0);
        \draw[obj3, ->-=.75] (0,0) -- (-\l,0);
        \draw[obj3, -<-=.47] (\r*\l+.5*\l,0) -- (\r*\sh*\l+\l,0);
        \node[mps] at (0,0) {};
        \node[mps] at (\r*\sh*\l,0) {};
        \node[xshift=15.2pt] at (0,0) {${\sss \varphi^{\alpha_1 \! \alpha_2}_{\alpha_3}}$};
        \node[xshift=-14pt] at (\r*\sh*\l,0) {${\sss \bar \varphi^{\alpha_1 \! \alpha_2}_{\alpha_4}}$};
    \end{tikzpicture}
}

\newcommand{\orthoFusionGen}[0]{
    \begin{tikzpicture}[scale=1,baseline={([yshift=-.55ex]current bounding box.center)}]
        \def\l{.6};
        \def\sh{1.23};
        \def\r{2.5};
        \draw[obj3, -<-=.46, rounded corners=2pt] (0,0) -- (0,\sh*\l/2) -- (\r*\sh*\l,\sh*\l/2) -- (\r*\sh*\l,0);
        \draw[obj3, -<-=.46, rounded corners=2pt] (0,0) -- (0,-\sh*\l/2) -- (\r*\sh*\l,-\sh*\l/2) -- (\r*\sh*\l,0);
        \draw[obj3, ->-=.75] (0,0) -- (-\l,0);
        \draw[obj3, -<-=.47] (\r*\l+.5*\l,0) -- (\r*\sh*\l+\l,0);
        \node[mps] at (0,0) {};
        \node[mps] at (\r*\sh*\l,0) {};
        \node[xshift=15.2pt] at (0,0) {${\sss \varphi^{V_1 \! V_2}_{V_3}}$};
        \node[xshift=-14pt] at (\r*\sh*\l,0) {${\sss \bar \varphi^{V_1 \! V_2}_{V_4}}$};
    \end{tikzpicture}
}

\newcommand{\orthoAction}[0]{
    \begin{tikzpicture}[scale=1,baseline={([yshift=-1.9ex]current bounding box.center)}]
        \def\l{.6};
        \def\sh{1.23};
        \draw[obj3, -<-=.48, rounded corners=2pt] (0,0) -- (0,\sh*\l) -- (2*\sh*\l,\sh*\l) -- (2*\sh*\l,0);
        \draw[obj2, -<-=.46, rounded corners=2pt] (0,0) -- (2*\sh*\l,0);
        \draw[obj2, ->-=.75] (0,0) -- (-\l,0);
        \draw[obj2, -<-=.47] (2*\sh*\l,0) -- (2*\sh*\l+\l,0);
        \node[mps] at (0,0) {};
        \node[mps] at (2*\sh*\l,0) {};
        \node[xshift=7pt, yshift=7pt] at (0,0) {${\sss \phi^{\alpha}}$};
        \node[xshift=-7pt, yshift=7pt] at (2*\sh*\l,0) {${\sss \bar \phi^{\alpha}}$};
        \node[below] at (\sh*\l,0) {${\sss \gamma}$};
    \end{tikzpicture}
}

\newcommand{\orthoActionGen}[0]{
    \begin{tikzpicture}[scale=1,baseline={([yshift=-2.87ex]current bounding box.center)}]
        \def\l{.6};
        \def\sh{1.23};
        \def\r{2.5};
        \draw[obj3, -<-=.48, rounded corners=2pt] (0,0) -- (0,\sh*\l) -- (\r*\sh*\l,\sh*\l) -- (\r*\sh*\l,0);
        \draw[obj2, -<-=.46, rounded corners=2pt] (0,0) -- (\r*\sh*\l,0);
        \draw[obj2, ->-=.75] (0,0) -- (-\l,0);
        \draw[obj2, -<-=.47] (\r*\sh*\l,0) -- (\r*\sh*\l+\l,0);
        \node[mps] at (0,0) {};
        \node[mps] at (\r*\sh*\l,0) {};
        \node[xshift=12pt, yshift=7pt] at (0,0) {${\sss \phi^{V \! M_1}_{M_2}}$};
        \node[xshift=-10pt, yshift=7pt] at (\r*\sh*\l,0) {${\sss \bar \phi^{V \! M_1}_{M_3}}$};
    \end{tikzpicture}
}

\newcommand{\fusionMPO}[4]{
    \begin{tikzpicture}[scale=1,baseline={([yshift=-.55ex]current bounding box.center)}]
        \def\l{.6};
        \def\sh{1.23};
        \ifthenelse{#1=1}{
            \draw[obj3, -<-=.47] (0,0) -- (\l,0);
            \draw[obj3, ->-=.75] (0,0) -- (-\l,0);
            \draw[obj3, -<-=.47] (0,\sh*\l) -- (\l,\sh*\l);
            \draw[obj3, ->-=.75] (0,\sh*\l) -- (-\l,\sh*\l);
            \draw[obj1, ->-=.75] (0,0) -- (0,\l);
            \draw[obj1, ->-=.75] (0,\sh*\l) -- (0,\sh*\l+\l);
            \draw[obj1, -<-=.47] (0,0) -- (0,-\l);
            \node[mpo] at (0,0) {};
            \node[mpo] at (0,\sh*\l) {};
            \node[xshift=10pt, yshift=6pt] at (0,0) {${\sss #3}$};
            \node[xshift=10pt, yshift=6pt] at (0,\sh*\l) {${\sss #2}$};
        }{}
        \ifthenelse{#1=2}{
            \draw[obj1, ->-=.75] (0,0) -- (0,\l);
            \draw[obj1, -<-=.47] (0,0) -- (0,-\l);
            \draw[obj3, -<-=.47] (0,0) -- (\l,0);
            \draw[obj3, ->-=.75] (0,0) -- (-\l,0);
            \draw[obj3, -<-=.62, rounded corners=2pt] (\sh*\l,0) -- (\sh*\l,\sh*\l/2) -- (\sh*\l+\l,\sh*\l/2);
            \draw[obj3, -<-=.62, rounded corners=2pt] (\sh*\l,0) -- (\sh*\l,-\sh*\l/2) -- (\sh*\l+\l,-\sh*\l/2);
            \draw[obj3, ->-=.78, rounded corners=2pt] (-\sh*\l,0) -- (-\sh*\l,\sh*\l/2) -- (-\sh*\l-\l,\sh*\l/2);
            \draw[obj3, ->-=.78, rounded corners=2pt] (-\sh*\l,0) -- (-\sh*\l,-\sh*\l/2) -- (-\sh*\l-\l,-\sh*\l/2);
            \node[mpo] at (0,0) {};
            \node[xshift=10pt, yshift=6pt] at (0,0) {${\sss #2}$};
            \node[mps] at (\sh*\l,0) {};
            \node[mps] at (-\sh*\l,0) {};
            \node[xshift=16pt] at (\sh*\l,0) {${\sss #3}$};
            \node[xshift=-15pt] at (-\sh*\l,0) {${\sss #4}$};
        }{}
    \end{tikzpicture}
}

\newcommand{\assocAction}[1]{
    \begin{tikzpicture}[scale=1,baseline={([yshift=-4.1ex]current bounding box.center)}]
        \def\l{.6};
        \def\sh{1.23};
        \ifthenelse{#1=1}{
            \draw[obj2, -<-=.47] (0,0) -- (\l,0);
            \draw[obj2, ->-=.75] (0,0) -- (-\l,0);
            \draw[obj2, -<-=.47] (\sh*\l,0) -- (2*\sh*\l+\l,0);
            \draw[obj2, ->-=.64] (-\l,0) -- (-2*\sh*\l-\l,0);
            \draw[obj3, -<-=.62, rounded corners=2pt] (2*\sh*\l,1.5*\sh*\l) -- (2*\sh*\l,2*\sh*\l) -- (2*\sh*\l+\l,2*\sh*\l);
            \draw[obj3, -<-=.62, rounded corners=2pt] (2*\sh*\l,1.5*\sh*\l) -- (2*\sh*\l,\sh*\l) -- (2*\sh*\l+\l,\sh*\l);
            \draw[obj3, ->-=.78, rounded corners=2pt] (-2*\sh*\l,1.5*\sh*\l) -- (-2*\sh*\l,\sh*\l) -- (-2*\sh*\l-\l,\sh*\l);
            \draw[obj3, ->-=.78, rounded corners=2pt] (-2*\sh*\l,1.5*\sh*\l) -- (-2*\sh*\l,2*\sh*\l) -- (-2*\sh*\l-\l,2*\sh*\l);
            \draw[obj3, -<-=.73, rounded corners=2pt] (\sh*\l,0) -- (\sh*\l,1.5*\sh*\l) -- (2*\sh*\l,1.5*\sh*\l);
            \draw[obj3, ->-=.82, rounded corners=2pt] (-\sh*\l,0) -- (-\sh*\l,1.5*\sh*\l) -- (-2*\sh*\l,1.5*\sh*\l);
            \draw[obj1, ->-=.7] (0,0) -- (0,\sh*\l);
            \node[mps] at (0,0) {};
            \node[mps] at (2*\sh*\l,1.5*\sh*\l) {};
            \node[mps] at (-2*\sh*\l,1.5*\sh*\l) {};
            \node[mps] at (\sh*\l,0) {};
            \node[mps] at (-\sh*\l,0) {};
            \node[yshift=-8pt] at (0,0) {${\sss \rho}$};
            \node[xshift=9pt, yshift=7pt] at (\sh*\l,0) {${\sss \phi^{\alpha_3}}$};
            \node[xshift=-7pt, yshift=7pt] at (-\sh*\l,0) {${\sss \bar \phi^{\alpha_3}}$};
            \node[xshift=16pt] at (2*\sh*\l,1.5*\sh*\l) {${\sss  \varphi^{\alpha_1 \! \alpha_2}_{\alpha_3}}$};
            \node[xshift=-15pt] at (-2*\sh*\l,1.5*\sh*\l) {${\sss \bar \varphi^{\alpha_1 \! \alpha_2}_{\alpha_3} }$};
            \node[yshift=-6pt] at (-2*\sh*\l,0) {${\sss \gamma_1}$};
            \node[yshift=-6pt] at (2*\sh*\l,0) {${\sss \gamma_1}$};
        }{}
        \ifthenelse{#1=2}{
            \draw[obj2, -<-=.47] (0,0) -- (\l,0);
            \draw[obj2, -<-=.47] (2*\sh*\l,0) -- (2*\sh*\l+\l,0);
            \draw[obj2, ->-=.75] (-2*\sh*\l,0) -- (-2*\sh*\l-\l,0);
            \draw[obj2, -<-=.47] (\sh*\l,0) -- (\sh*\l+\l,0);
            \draw[obj2, ->-=.75] (0,0) -- (-\l,0);
            \draw[obj2, ->-=.75] (-\sh*\l,0) -- (-\sh*\l-\l,0);
            \draw[obj3, -<-=.72, rounded corners=2pt] (2*\sh*\l,0) -- (2*\sh*\l,\sh*\l) -- (2*\sh*\l+\l,\sh*\l);
            \draw[obj3, ->-=.85, rounded corners=2pt] (-2*\sh*\l,0) -- (-2*\sh*\l,\sh*\l) -- (-2*\sh*\l-\l,\sh*\l);
            \draw[obj3, -<-=.2, rounded corners=2pt] (-2*\sh*\l-\l,2*\sh*\l) -- (-\sh*\l,2*\sh*\l) -- (-\sh*\l,0);
            \draw[obj3, ->-=.26, rounded corners=2pt] (2*\sh*\l+\l,2*\sh*\l) -- (\sh*\l,2*\sh*\l) -- (\sh*\l,0);
            \draw[obj1, ->-=.7] (0,0) -- (0,\sh*\l);
            \node[mps] at (0,0) {};
            \node[mps] at (\sh*\l,0) {};
            \node[mps] at (-\sh*\l,0) {};
            \node[mps] at (2*\sh*\l,0) {};
            \node[mps] at (-2*\sh*\l,0) {};
            \node[yshift=-8pt] at (0,0) {${\sss \rho}$};
            \node[xshift=9pt, yshift=7pt] at (\sh*\l,0) {${\sss \phi^{\alpha_1}}$};
            \node[xshift=-7pt, yshift=7pt] at (-\sh*\l,0) {${\sss \bar \phi^{\alpha_1} \,}$};
            \node[xshift=9pt, yshift=7pt] at (2*\sh*\l,0) {${\sss  \phi^{\alpha_2}}$};
            \node[xshift=-7pt, yshift=7pt] at (-2*\sh*\l,0) {${\sss \bar \phi^{\alpha_2} }$};
            \node[yshift=-6pt] at (-2*\sh*\l-\sh*\l/2,0) {${\sss \gamma_1}$};
            \node[yshift=-6pt] at (2*\sh*\l+\sh*\l/2,0) {${\sss \gamma_1}$};
            \node[yshift=-6pt] at (-\sh*\l-\sh*\l/2,0) {${\sss \gamma_3}$};
            \node[yshift=-6pt] at (\sh*\l+\sh*\l/2,0) {${\sss \gamma_3}$};
        }{}
    \end{tikzpicture}
}

\newcommand{\moduleAssoc}[1]{
    \begin{tikzpicture}[scale=1,baseline={([yshift=-4.25
    ex]current bounding box.center)}]
        \def\l{.6};
        \def\sh{1.23};
        \ifthenelse{#1=1}{
            \draw[obj2, ->-=.75] (0,0) -- (-\l,0);
            \draw[obj2, -<-=.46] (0,0) -- (\sh*\l+\l,0);
            \draw[obj3, -<-=.62, rounded corners=2pt] (\sh*\l,1.5*\sh*\l) -- (\sh*\l,2*\sh*\l) -- (\sh*\l+\l,2*\sh*\l);
            \draw[obj3, -<-=.62, rounded corners=2pt] (\sh*\l,1.5*\sh*\l) -- (\sh*\l,\sh*\l) -- (\sh*\l+\l,\sh*\l);
            \draw[obj3, -<-=.73, rounded corners=2pt] (0,0) -- (0,1.5*\sh*\l) -- (\sh*\l,1.5*\sh*\l);
            \node[mps] at (\sh*\l,1.5*\sh*\l) {};
            \node[mps] at (0,0) {};
            \node[xshift=9pt, yshift=7pt] at (0,0) {${\sss \phi^{\alpha_3}}$};
            \node[xshift=16pt] at (\sh*\l,1.5*\sh*\l) {${\sss \varphi^{\alpha_1 \! \alpha_2}_{\alpha_3}}$};
            \node[yshift=-6pt] at (-\sh*\l/2,0) {${\sss \gamma_2}$};
            \node[yshift=-6pt] at (\sh*\l,0) {${\sss \gamma_1}$};
        }{}
        \ifthenelse{#1=2}{
            \draw[obj2, ->-=.75] (0,0) -- (-\l,0);
            \draw[obj2, -<-=.47] (0,0) -- (\l,0);
            \draw[obj2, -<-=.47] (\sh*\l,0) -- (\sh*\l+\l,0);
            \draw[obj3, -<-=.72, rounded corners=2pt] (\sh*\l,0) -- (\sh*\l,\sh*\l) -- (\sh*\l+\l,\sh*\l);
            \draw[obj3, ->-=.26, rounded corners=2pt] (\sh*\l+\l,2*\sh*\l) -- (0,2*\sh*\l) -- (0,0);
            \node[mps] at (0,0) {};
            \node[mps] at (\sh*\l,0) {};
            \node[xshift=9pt, yshift=7pt] at (0,0) {${\sss \, \phi^{\alpha_1}}$};
            \node[xshift=9pt, yshift=7pt] at (\sh*\l,0) {${\sss  \phi^{\alpha_2}}$};
            \node[yshift=-6pt] at (-\sh*\l/2,0) {${\sss \gamma_2}$};
            \node[yshift=-6pt] at (\sh*\l/2,0) {${\sss \gamma_3}$};
            \node[yshift=-6pt] at (\sh*\l+\sh*\l/2,0) {${\sss \gamma_1}$};
        }{}
    \end{tikzpicture}
}

\newcommand{\Fsym}[0]{
    \begin{tikzpicture}[scale=1,baseline={([yshift=-4.25
    ex]current bounding box.center)}]
        \def\l{.6};
        \def\sh{1.23};
        \draw[obj2, ->-=.75] (0,0) -- (-\l,0);
        \draw[obj2, -<-=.46] (0,0) -- (\sh*\l+2*\l,0);
        \draw[obj2, -<-=.47] (\sh*\l+2*\l,0) -- (\sh*\l+3*\l,0);
        \draw[obj2, -<-=.47] (2*\sh*\l+2*\l,0) -- (2*\sh*\l+3*\l,0);
        \draw[obj3, -<-=.33, rounded corners=2pt] (\sh*\l,1.5*\sh*\l) -- (\sh*\l,2*\sh*\l) -- (2*\sh*\l+2*\l,2*\sh*\l) -- (2*\sh*\l+2*\l,0);
        \draw[obj3, -<-=.4, rounded corners=2pt] (\sh*\l,1.5*\sh*\l) -- (\sh*\l,\sh*\l) -- (\sh*\l+2*\l,\sh*\l) -- (\sh*\l+2*\l,0);
        \draw[obj3, -<-=.73, rounded corners=2pt] (0,0) -- (0,1.5*\sh*\l) -- (\sh*\l,1.5*\sh*\l);
        \node[mps] at (2*\sh*\l+2*\l,0) {};
        \node[mps] at (\sh*\l+2*\l,0) {};
        \node[mps] at (\sh*\l,1.5*\sh*\l) {};
        \node[mps] at (0,0) {};
        \node[xshift=9pt, yshift=7pt] at (0,0) {${\sss \phi^{\alpha_3}}$};
        \node[xshift=16pt] at (\sh*\l,1.5*\sh*\l) {${\sss \varphi^{\alpha_1 \! \alpha_2}_{\alpha_3}}$};
        \node[yshift=-6pt] at (-\sh*\l/2,0) {${\sss \gamma_2}$};
        \node[yshift=-6pt] at (\sh*\l/2+\l,0) {${\sss \gamma_1}$};
        \node[yshift=-6pt] at (3*\sh*\l/2+2*\l,0) {${\sss \gamma_3}$}; 
        \node[yshift=-6pt] at (2*\sh*\l+2*\l+\sh*\l/2,0) {${\sss \gamma_2}$};
        \node[xshift=-7pt, yshift=7pt] at (\sh*\l+2*\l,0) {${\sss \bar \phi^{\alpha_2}}$};
        \node[xshift=-7pt, yshift=7pt] at (2*\sh*\l+2*\l,0) {${\sss \bar \phi^{\alpha_1}}$}; 
    \end{tikzpicture}
}


\title{A non-semisimple non-invertible symmetry}

\author{\textsc{Clement Delcamp}}
\email{delcamp@ihes.fr}
\affiliation{Institut des Hautes \'Etudes Scientifiques, Bures-sur-Yvette, France}
\author{\textsc{Edmund Heng}}
\email{heng@ihes.fr}
\affiliation{Institut des Hautes \'Etudes Scientifiques, Bures-sur-Yvette, France}
\author{\textsc{Matthew Yu}}
\email{yum@maths.ox.ac.uk}
\affiliation{Mathematical Institute, University of Oxford, Woodstock Road, Oxford, UK}

\begin{abstract}
\noindent
We investigate the action of a non-invertible symmetry on spins chains whose topological lines are labelled by representations of the four-dimensional Taft algebra. The main peculiarity of this symmetry is the existence of junctions between distinct indecomposable lines. Sacrificing Hermiticity, we construct several symmetric, frustration-free, gapped Hamiltonians with real spectra and analyse their ground state subspaces. Our study reveals two intriguing phenomena. First, we identify a smooth path of gapped symmetric Hamiltonians whose ground states transform inequivalently under the symmetry. Second, we find a model where a product state and the so-called W state spontaneously break the symmetry, and propose an explanation for the indistinguishability of these two states in the infinite-volume limit in terms of the symmetry category. 
\end{abstract}

\maketitle

\noindent
{\bf Introduction:} Defining internal symmetry in a quantum theory through the lens of \emph{topological defects} has opened the door to generalised notions of symmetry, including some arising from \emph{non-invertible} transformations \cite{Gaiotto:2014kfa,Freed:2022qnc}. 
Mathematically, it is understood that in (1+1)d the framework of \emph{fusion category} theory offers an axiomatisation for \emph{finite} non-invertible symmetries, extending the group theoretic framework of ordinary symmetries \cite{Frohlich:2006ch,Bhardwaj:2017xup,Chang:2018iay,Thorngren:2019iar,PhysRevResearch.2.043086,Schafer-Nameki:2023jdn,Shao:2023gho}. Importantly, such fusion categories are \emph{semisimple}, which physically ensures that no local operators can transform one indecomposable topological line defect into a distinct one. Given such a symmetry, a classification of (bosonic) symmetric gapped phases has been proposed, extending the ordinary Landau paradigm \cite{Thorngren:2019iar,Komargodski:2020mxz}. 
In particular, for each gapped phase, a commuting projector Hamiltonian representing the corresponding gapped phase can be explicitly constructed within the \emph{anyonic chain} framework \cite{Feiguin:2006ydp,PhysRevB.87.235120,Buican:2017rxc,Inamura:2021szw,Lootens:2021tet,Lootens:2022avn,Bhardwaj:2024kvy}. 
Moreover, both ground states and symmetry operators can be efficiently parametrised in terms of \emph{tensor networks} \cite{Fannes1992,Perez-Garcia:2006nqo,Lootens:2021tet,Lootens:2022avn,Lootens:2024gfp,Garre-Rubio:2022uum,Molnar:2022nmh}.

What happens to these results when the symmetry structure is no longer required to be semisimple? Do new features arise in such cases? Using the tools of \cite{Lootens:2021tet,Inamura:2021szw,Lootens:2022avn,Garre-Rubio:2022uum,Bhardwaj:2024kvy}, we explore these questions through investigating a specific example: a symmetry encoded into the category of modules over the Taft algebra of dimension 4. This \emph{non-semisimple tensor category} describes topological line defects that are comprised of simpler line defects, and yet cannot be decomposed as a direct sum of them, implying the existence of local operators transforming distinct line defects into one another. Notably, gapped Hamiltonians with such a symmetry are generally not self-adjoint. Nonetheless, this does not preclude the possibility of finding Hamiltonians with a \emph{real spectrum}. 

Our study highlights two phenomena: On the one hand, we find a smooth $\mathbb{S}^1$-parametrised path of gapped symmetric Hamiltonians---which we would interpret as representing the same gapped phase by extending the usual definition to include non-self-adjoint Hamiltonians---yet whose ground states transform inequivalently under the non-semisimple symmetry, in the sense of ref.~\cite{Garre-Rubio:2022uum}. On the other hand, we construct a Hamiltonian whose two degenerate ground states spontaneously break the non-semisimple symmetry. Moreover, they are indistinguishable in the infinite volume limit, and thus provide a unique vacuum. We relate this indistinguishability to the existence of maps between the objects in a category that are respectively associated with the two ground states.

Although mathematically ubiquitous, non-semisimple categories have not received widespread attention in physics yet.   
They have primarily seen applications in the context of \emph{non-rational conformal field theories}  \cite{Kausch:2000fu,Gaberdiel:2001tr,Fuchs:2006nx,Fuchs:2013lda,Creutzig:2013hma}, and lattice regularisations thereof \cite{Pasquier:1989kd,saleur2006enlarged,Read:2007qq,Read:2007ti,Gainutdinov:2012qy,Belletete_2017}, as well as twisted \emph{supersymmetric topological field theories} \cite{Creutzig:2021ext,Garner:2022rwe,ArabiArdehali:2024vli}. Recently, there has been a lot of progress in constructing three-dimensional state-sum invariants from certain non-semisimple categories \cite{Geer_2022,Geer:2022qoa,costantino,Iulianelli:2024icl,Hofer:2024cnu}, which we expect to be able to relate to our work through the scope of the \emph{symmetry topological field theory} construction, see e.g. \cite{KONG201762,PhysRevResearch.2.043086,PhysRevResearch.2.033417,Ji:2021esj,Chatterjee:2022kxb,Freed:2022qnc,Huang:2023pyk,Huang:2024ror,PhysRevLett.112.247202,PhysRevLett.121.177203,Apruzzi:2021nmk,Kaidi:2022cpf,Kaidi:2023maf,Delcamp:2024cfp}.

\medskip \noindent
Note that the exposition in the main text is self-contained and all the results can be verified using the notions introduced there. Nonetheless, our manuscript is complemented by an appendix that compiles various mathematical constructions and derivations, which both motivate our study, and shed light on the results presented in the main text.  Although the focus remains on the category of modules over the Taft algebra of dimension 4, we expect the formalism developed in this appendix to be relevant, more generally, for any non-anomalous non-semisimple non-invertible symmetry admitting finitely-many indecomposable lines.

\bigskip
\noindent
{\bf Spontaneous breaking of the invertible symmetry:} We begin with a study of one-dimensional quantum lattice models with open boundary conditions representing gapped phases spontaneously breaking a $\mathbb Z/2\mathbb Z$ symmetry.

Let $\Lambda$ be a finite subset of the lattice $\mathbb Z$. To each element $\msf i \in \Lambda$, hereafter referred to as a `site', we assign a copy of the algebra $\msf{Mat}_\mathbb C(2)$ of $2 \times 2$ matrices with complex numbers. We identify $\bigotimes_{\msf i \in \Lambda} \msf{Mat}_\mathbb C(2)_{\! \{\msf i\}}$ with the algebra of (bounded) operators acting on the `microscopic' Hilbert space $\mc H_\Lambda = \bigotimes_{\msf i \in \Lambda} \mathbb C^2$ of `spin' degrees of freedom on $\Lambda$. Throughout this manuscript, we work in the computational  basis $\mathbb C^2 \cong \mathbb C\{|0\ra,|1\ra\}$. As is customary, for any $\msf i \in \Lambda$, $O_\msf i$ will denote the embedding of $O \in \msf{Mat}_\mathbb C(2)_{\! \{\msf i\}}$ into $\bigotimes_{\msf i \in \Lambda}\msf{Mat}_\mathbb C(2)_{\!\{\msf i\}}$ by tensoring with the identity matrix.

Given $\xi \in \rU(1)$, suppose the dynamics of the spin degrees of freedom is governed by the nearest-neighbour Hamiltonian operator $\mathbb H(\xi)_\Lambda = - \sum_{\msf i \in \Lambda} \mathbb h(\xi)_{\msf i,\msf i+1} :\mc H_\Lambda \to \mc H_\Lambda$ defined in terms of local commuting projectors
\begin{equation}
    \label{eq:localOpXX}
    \mathbb h(\xi)_{\msf i,\msf i+1} := \frac{1}{2}\big[\mathbb I_\msf i \otimes \mathbb I_{\msf i+1} + \sigma^x(\xi)_\msf i \otimes \sigma^x(\xi)_{\msf i+1} \big] \, ,
\end{equation}
where $\mathbb I$ is the identity matrix and $\sigma^x(\xi) := \sqrt \xi \begin{psmallmatrix} 0 & 1 / \xi \\ 1 & 0\end{psmallmatrix}$. The hamiltonian $\mathbb H(\xi)_\Lambda$ is gapped, possesses a $\mathbb Z/2\mathbb Z$ symmetry generated by $\prod_{\msf i \in \Lambda}\sigma_\msf i^z$ with $\sigma^z = \begin{psmallmatrix}1 & 0 \\ 0 & -1\end{psmallmatrix}$, and its two-dimensional ground state subspace is spanned by tensor product states $|+\xi \ra^{\otimes | \Lambda |}$ and $|-\xi \ra^{\otimes | \Lambda |}$, where $| \pm \xi \ra := | 0 \ra \pm \sqrt \xi |1 \ra$. Since $|+\xi \ra^{\otimes |\Lambda|}$ and $|-\xi \ra^{\otimes |\Lambda|}$ are mapped onto each other under the action of $\prod_{\msf i \in \Lambda} \sigma^z_\msf i$, the symmetry $\mathbb Z / 2 \mathbb Z$ of $\mathbb H(\xi)_\Lambda$ is spontaneously broken in its ground state subspace. Furthermore, it is clear that the whole $\mathbb S^1$-parametrised family of Hamiltonians $\{ \mathbb H(\xi)_\Lambda \, | \, \xi \in \rU(1)\}$ represents the same spontaneously symmetry broken gapped phase with respect to the $\mathbb Z /2 \mathbb Z$ symmetry generated by $\prod_{\msf i \in \Lambda}\sigma^z_\msf i$.

Given $\xi \in \rU(1)$, suppose instead that the dynamics is governed by the nearest neighbour Hamiltonian operator $\widetilde{\mathbb H}(\xi)_\Lambda = -\sum_{\msf i \in \Lambda} \widetilde{\mathbb h}(\xi)_{\msf i, \msf i+1}$ defined in terms of local projectors (see app.~\ref{app:moduleCat} for motivation)
\begin{equation}
    \label{eq:localOpXM}
    \widetilde{\mathbb h}(\xi)_{\msf i, \msf i+1} := \mathbb I_\msf i \otimes (\sigma^+ \sigma^-)_{\msf i+1} + \sqrt{\xi} \, \sigma^x(\xi)_\msf i \otimes \sigma^-_{\msf i+1}  \, ,
\end{equation}
where $\sigma^+  := \begin{psmallmatrix} 0 & 1 \\ 0  & 0 \end{psmallmatrix}$ and $\sigma^-  := \begin{psmallmatrix} 0 & 0 \\ 1  & 0 \end{psmallmatrix}$. Let us immediately point out the obvious fact that this Hamiltonian is \emph{non-hermitian}. Nevertheless, its eigenvalues can be verified to be all real negative. Moreover, in spite of the local projectors not commuting with one another, this alternative Hamiltonian is \emph{frustration-free} since the ground states minimise each local term individually. Additionally, it showcases a spectral gap.\footnote{Performing a Jordan--Wigner transformation, one can readily verify that the spectrum is valued in $\mathbb Z_{\leq 0}$ and that the degeneracy of eigenvalue $-k$ is $2{|\Lambda|-1 \choose k}$.}
Finally, it possesses the same $\mathbb Z/2\mathbb Z$ symmetry as $\mathbb H(\xi)_\Lambda$, and its two-dimensional ground state subspace is also spanned by the tensor product states $|+\xi \ra^{\otimes |\Lambda |}$ and $|-\xi \ra^{\otimes | \Lambda |}$. Thus, for each $\xi \in \rU(1)$, the symmetry $\mathbb Z / 2 \mathbb Z$ is again spontaneously broken in the ground state subspace of $\widetilde{\mathbb H}(\xi)_\Lambda$. But, Hamiltonian $\widetilde{\mathbb H}(\xi)_\Lambda$ happens to have a richer symmetry structure.  

\medskip
\noindent
Let us reveal the existence of additional symmetry operators. Let $\omega_0$ be a collection of linear maps
$(\omega_0)^{d_2}_{d_1} : \mathbb C^2 \to \mathbb C^2$, with $d_1,d_2 \in \{0,1\}$, best defined graphically via the following tensor
\begin{equation}
\begin{split}
    \label{eq:MPO_P0}
    \MPOT{\omega_0}{}{}{}{} \!\!\!\!
    &\equiv \!\!\! \sum_{\substack{d_1,d_2 \in \{0,1\} \\ b_1,b_2 \in \{0,1\} }} \!\!\!\!\!
    \MPOT{\omega_0}{b_2}{\, d_1}{d_2 \,}{b_1} \, 
    | b_2 \ra \la b_1 |  \otimes | d_2 \ra \la d_1 |
\end{split}
\end{equation}
such that (see app.~\ref{app:chain} for motivation) 
\begin{equation*}
\begin{split}
    &(\omega_0)_0^0 \equiv \MPOT{\omega_0}{}{0}{0}{} :=  \mathbb I
    \, , \q 
    (\omega_0)_0^1 \equiv  \MPOT{\omega_0}{}{0}{1}{} := \sigma^- \! ,
    \\[-.5em]
    & (\omega_0)_1^0 \equiv \MPOT{\omega_0}{}{1}{0}{} := 0 \,
     , \q 
    (\omega_0)_1^1 \equiv \MPOT{\omega_0}{}{1}{1}{} := \sigma^z .
\end{split}
\end{equation*}
Then, define the collection $\widehat \omega_0$ of matrix product operators $ (\widehat \omega_0)_{d_1}^{d_{|\Lambda|+1}} \equiv \la d_{|\Lambda|+1} | \, \widehat \omega_0 \, | d_1 \ra : \mc H_\Lambda \to \mc H_\Lambda$ with open boundary conditions $|d_1 \ra, |d_{|\Lambda|+1} \ra \in \mathbb C^2$ via
\begin{align}  
    \label{eq:MPO}
    (\widehat \omega_0)_{d_1}^{d_{|\Lambda|+1}} \! &:= \!\!\!\! \sum_{d_2,\ldots,d_{|\Lambda|}} \!\!\!
    (\omega_0)^{d_{|\Lambda|+1}}_{d_{|\Lambda|}} \otimes \cdots \otimes (\omega_0)^{d_3}_{d_2} \otimes (\omega_0)^{d_2}_{d_1} 
    \\[-.2em] \nn
    & \equiv \la d_{|\Lambda|+1} |\,  \MPO{\omega_0} \, | d_1 \ra \, .
\end{align}
Notice that $(\widehat \omega_0)_0^0$ and $(\widehat \omega_0)_1^0$ coincide with the identity and the zero operators, respectively, while $(\widehat \omega_0)^1_1$ gives the generator of the $\mathbb Z / 2 \mathbb Z$ symmetry. As such, these three operators commute with the Hamiltonian $\widetilde{\mathbb H}(\xi)_{\Lambda}$. It follows from
\begin{equation}
    \widetilde{\mathbb h}(\xi) \cdot \big[\sigma^- \otimes \mathbb I + \sigma^z \otimes \sigma^- \big]
    =  \big[\sigma^- \otimes \mathbb I + \sigma^z \otimes \sigma^- \big] \cdot \widetilde{\mathbb h}(\xi) \, 
\end{equation}
that the matrix product operator $(\widehat \omega_0)^1_0$ also commutes with the Hamiltonian. 
Moreover, any composition of these operators results in another operator commuting with the Hamiltonian. In particular, we denote by $\widehat \omega_1$ the collection of operators defined in terms of the tensor
\begin{equation}
    \MPOT{\omega_1}{}{}{}{} \!\!\! := \!\!\! \MPOTSpe{\omega_0}{}{}{}{} \!\! ,
\end{equation}
which is obtained by precomposing the collection of operators $\widehat \omega_0$ with $\prod_{\msf i \in \Lambda}\sigma^z_\msf i$. Furthermore, notice that we have $((\widehat \omega_0)^1_0)^2 = 0$, which implies in particular that $(\widehat \omega_0)^1_0$ is non-invertible. Crucially, if we were to work with closed boundary conditions instead, the collection $\widehat \omega_0$ of operators would boil down to the single operator $(\widehat \omega_0)_0^0 + (\widehat \omega_0)_1^1$, and would thus be redundant with respect to the $\mathbb Z / 2 \mathbb Z$ symmetry. 

In order for the collections $\widehat{\omega}_\alpha$, with $\alpha \in \{0,1\}$, of matrix product operators to define a symmetry, one further requires the existence of junctions of symmetry operators which themselves host vector spaces of local operators. These are provided by linear maps $\varphi^{\alpha_1 \!  \alpha_2}_{\alpha_3} : \mathbb C^2 \otimes \mathbb C^2 \to \mathbb C^2$, with $\alpha_1,\alpha_2,\alpha_3 \in \{0,1\}$, defined graphically as
\begin{equation}
    \fusionT{\;\;\;\, \varphi_{\alpha_3}^{\alpha_1 \! \alpha_2}}{}{}{} 
    \!\! \equiv \! 
    \sum_{d_1,d_2,d_3} \!\! \fusionT{\;\;\;\, \varphi_{\alpha_3}^{\alpha_1 \! \alpha_2}}{\, d_1}{\, d_2}{d_3} \, |d_3 \ra \la d_1,d_2 | \, ,
\end{equation}
for which we list below the non-vanishing entries for $\alpha_1=\alpha_2$ (see app.~\ref{app:TC} for motivation):
\begin{align}
    \label{eq:fusion_P0}
    \fusionT{\varphi^{00}_0}{0}{0}{0} &=
    \fusionT{\varphi^{00}_0}{0}{1}{1} = 1 \, ,
    \\ \nn
    (-1)  \fusionT{\varphi^{00}_1}{1}{0}{0} &= 
    \fusionT{\varphi^{00}_1}{0}{1}{0} = \fusionT{\varphi^{00}_1}{1}{1}{1} = - 1 \, ,
    \\ \label{eq:fusion_P1}
    (-1) \fusionT{\varphi^{11}_0}{0}{1}{1} &=
    \fusionT{\varphi^{11}_0}{0}{0}{0} = 1 \, ,
    \\ \nn
    \fusionT{\varphi^{11}_1}{0}{1}{0} &= 
    \fusionT{\varphi^{11}_1}{1}{0}{0} = \fusionT{\varphi^{11}_1}{1}{1}{1} = 1 \, .
\end{align}
Together with linear maps $\bar \varphi^{\alpha_1 \! \alpha_2}_{\alpha_3} : \mathbb C^2 \to \mathbb C^2 \otimes \mathbb C^2$, with $\alpha_1,\alpha_2,\alpha_3 \in \{0,1\}$, verifying
\begin{equation}
    \label{eq:orthoFusion}
    \orthoFusion = \delta_{\alpha_3,\alpha_4} \,  \mathbb I_{\mathbb C^2} \, ,
\end{equation}
for every $\alpha_1, \alpha_2, \alpha_3,\alpha_4 \in \{0,1\}$, these allow us to locally fuse the matrix product operators defined in eq.~\eqref{eq:MPO} according to
\begin{equation}
    \fusionMPO{1}{\omega_{\alpha_1}}{\omega_{\alpha_2}}{} = \sum_{\alpha_3 \in \{0,1\}} \fusionMPO{2}{\omega_{\alpha_3}}{\varphi^{\alpha_1 \! \alpha_2}_{\alpha_3}}{\bar \varphi^{\alpha_1 \! \alpha_2}_{\alpha_3}} \! .
\end{equation}
In the appendices, we demonstrate that this fusion pattern is specific to a symmetry structure encoded into the  \emph{non-semisimple} tensor category $\Mod(\mc T_4)$ of modules over the Taft algebra $\mc T_4$ of dimension 4 (see app.~\ref{app:Taft}). 

\medskip
\noindent
We claim that, for any two distinct values of the parameter $\xi \in \rU(1)$, states in the ground state subspace of the Hamiltonian $\widetilde{\mathbb H}(\xi)_\Lambda$ transform inequivalently under the $\Mod(\mc T_4)$ symmetry in the sense of ref.~\cite{Garre-Rubio:2022uum}.

Consider an arbitrary state in the ground state subspace spanned by $|+\xi \ra^{\otimes |\Lambda|}$ and $|-\xi \ra^{\otimes |\Lambda|}$. In general, such an arbitrary state is not a tensor product state, and it is best expressed as a matrix product state. We do so in the following way. 
Let $\rho$ be a collection of vectors
$\rho_{\gamma_1}^{\gamma_2} \in  \mathbb C^2$, with $\gamma_1,\gamma_2 \in \{0,1\}$, defined graphically via the following tensor\footnote{As suggested by the notation, indices $\gamma_1$ and $\gamma_2$ are not quite on the same footing as $b$. One should think of the former as labelling one-dimensional blocks rather than basis vectors.}
\begin{align}
    \label{eq:MPSXM}
    \MPST{\rho}{}{}{}{}{} \!\!\! \equiv \!\!\!
    \sum_{\substack{b \in \{0,1\} \\ \gamma_2,\gamma_1 \in \{0,1\}}} \hspace{-1.2em} \MPST{\rho}{b}{\gamma_2}{\gamma_1}{}{} \! | b \ra \otimes |\gamma_2 \ra \la \gamma_1 | 
\end{align}
such that (see app.~\ref{app:moduleCat} for motivation)
\begin{align*}
    &\rho^0_0 \equiv \!\!\!\! \MPST{\rho}{}{0}{0}{}{} \!\!\!\! :=  |+\xi \ra
    \, , \q 
    \rho^1_0 \equiv \!\!\!\! \MPST{\rho}{}{1}{0}{}{} \!\!\!\! := 0 \, ,
    \\ \nn
    & \rho^0_1 \equiv \!\!\!\! \MPST{\rho}{}{0}{1}{}{} \!\!\!\! := 0 \,
     , \q 
    \rho_1^1 \equiv \!\!\!\! \MPST{\rho}{}{1}{1}{}{} \!\!\!\! := |-\xi \ra \, .
\end{align*}
Given basis vectors $|\gamma_1 \ra, |\gamma_{|\Lambda|+1}\ra \in \mathbb C \oplus \mathbb C$ encoding a choice of open boundary conditions, we construct the corresponding ground state as
\begin{align}
    \label{eq:MPS}
    &\sum_{\gamma_2,\ldots,\gamma_{|\Lambda|}} \rho_{\gamma_{|\Lambda|}}^{\gamma_{|\Lambda|+1}} \otimes \cdots \otimes \rho_{\gamma_2}^{\gamma_3} \otimes \rho_{\gamma_1}^{\gamma_2}
    \\[-.2em] \nn 
    &\q \equiv \la \gamma_{|\Lambda|+1} \! | \raisebox{1.7pt}{\MPS{\rho}} \!  | \gamma_1 \ra \, .
\end{align}
Any state in the ground state subspace of $\mc H_\Lambda$ can be obtained via an appropriate linear combination of  open boundary conditions. Now, consider acting on such a state with the collections $\widehat \omega_\alpha$, with $\alpha \in \{0,1\}$, of matrix product operators defined in eq.~\eqref{eq:MPO}. The fact that the $\Mod(\mc T_4)$ symmetry preserves the ground state subspace implies the existence of linear maps $\phi^{\alpha} : \mathbb C^2 \otimes (\mathbb C \oplus \mathbb C) \to \mathbb C \oplus \mathbb C$, with $\alpha \in \{0,1\}$, defined graphically as
\begin{align}
    \label{eq:actionT}
    \actionT{\phi^{\alpha}}{}{}{} \!\!\!\! \equiv \sum_{\substack{d \in \{0,1\} \\ \gamma_1,\gamma_2 \in \{0,1\}}} \!\!\!\! \actionT{\phi^\alpha}{d}{\gamma_2}{\gamma_1} |\gamma_2 \ra \la d,\gamma_1 | \, ,
\end{align}
whose non-vanishing entries are given by (see app.~\ref{app:moduleCat} for motivation)
\begin{align*}
    \actionT{\phi^{0}}{0}{0}{0} \!\!\!\! &=  
    \, \actionT{\phi^{0}}{1}{1}{0} \!\!\!\! =  
    (-2 \sqrt \xi) \; \actionT{\phi^{0}}{0}{1}{0} \!\!\!\! = 1 \, ,
    \\ 
    \actionT{\phi^{0}}{1}{0}{1} \!\!\!\! &= \, 
    \actionT{\phi^{0}}{0}{1}{1} \!\!\!\! = 
    (2 \sqrt \xi) \; \actionT{\phi^{0}}{0}{0}{1} \!\!\!\! = 1 \, ,
\end{align*}
and
\begin{align*}
    \actionT{\phi^{1}}{1}{0}{0} \!\!\!\! &=  
    \, \actionT{\phi^{1}}{0}{1}{0} \!\!\!\! =  
    \, (2 \sqrt \xi) \; \actionT{\phi^{1}}{0}{0}{0} \!\!\!\! = 1 \, ,
    \\ \nn 
    \actionT{\phi^{1}}{1}{1}{1} \!\!\!\! &= \, 
    \actionT{\phi^{1}}{0}{0}{1} \!\!\!\! = 
    \, (-2 \sqrt \xi) \; \actionT{\phi^{1}}{0}{1}{1} \!\!\!\! = 1 \, .
\end{align*}
Together with linear maps $\bar \phi^{\alpha} : \mathbb C \oplus \mathbb C \to \mathbb C^2 \otimes (\mathbb C \oplus \mathbb C)$, with $\alpha \in \{0,1\}$, satisfying
\begin{align}
    \label{eq:orthoAction}
    \orthoAction =  
    \mathbb I_{\mathbb C \oplus \mathbb C} \, ,
\end{align}
for every $\gamma \in \{0,1\}$ and $\alpha \in \{0,1\}$, these allow us to compute the local action of the collections of symmetry operators $\widehat \omega_\alpha$, with $\alpha \in \{0,1\}$, on the ground state subspace according to
\begin{equation}
    \label{eq:actionMPO}
    \raisebox{12.5pt}{\actionMPO{1}{\rho}{\omega_\alpha}{}{}{\gamma} } \!\! = \raisebox{4.8pt}{\actionMPO{2}{\rho}{}{\phi^\alpha}{\bar \phi^\alpha}{\gamma} } \! ,
\end{equation}
for every $\gamma \in \{0,1\}$.
Combining eq.~\eqref{eq:orthoAction} and eq.~\eqref{eq:actionMPO}, one recovers in particular  $(\widehat \omega_0)_1^1 | \pm \xi \ra^{\otimes |\Lambda|} = | \mp \xi \ra^{\otimes |\Lambda|}$ and $(\widehat \omega_0)_0^1 | \pm \xi \ra^{\otimes |\Lambda|} = | + \xi \ra^{\otimes |\Lambda|} - | - \xi \ra^{\otimes |\Lambda|}$.

Now, consider the successive actions of two symmetry operators. One can explicitly verify the following associativity condition:
\begin{equation}
\begin{split}
    \label{eq:assocAction}
    &\sum_{\alpha_3 \in \{0,1\}} \!
     \!\!\!\!\!\!\! \assocAction{1} 
    \\  
    & \q = \!\! \sum_{\gamma_3 \in \{0,1\}} \!\! \assocAction{2} \, ,
\end{split}
\end{equation}
for every $\gamma_1 \in \{0,1\}$.
From orthogonality conditions \eqref{eq:orthoFusion} and \eqref{eq:orthoAction} follows the existence of so-called $\F{\act}$-symbols $\big(\F{\act}^{\alpha_1 \alpha_2 \gamma_1}_{\gamma_2} \big)_{\alpha_3}^{\gamma_3} \in \mathbb C$, for every $\gamma_1, \gamma_2, \gamma_3,\alpha_1,\alpha_2,\alpha_3 \in \{0,1\}$, satisfying
\begin{equation*}
    \moduleAssoc{1} \!\!\!\!\!  = \!\! \sum_{\gamma_3 \in \{0,1\}} \!\! \big(\F{\act}^{\alpha_1 \alpha_2 \gamma_1}_{\gamma_2} \big)_{\alpha_3}^{\gamma_3} \; \moduleAssoc{2} \, .
\end{equation*}
Explicitly, the $\F{\act}$-symbols evaluate to
\begin{equation}
    \big(\F{\act}^{\alpha_1 \alpha_2 \gamma_1}_{\gamma_2} \big)_{\alpha_3}^{\gamma_3}
    \, \mathbb I_\mathbb C = \Fsym \, .
\end{equation}
These $\F{\act}$-symbols can be verified to satisfy \emph{pentagon equations} involving so-called $F$-symbols, which can be constructed similarly in terms of junctions of symmetry operators only (see app.~\ref{app:moduleCat}).
We organise some of these symbols into the following matrices:
\begin{equation*}
\begin{split}
    \big(\F{\act}^{000}_0 \big)_{\alpha_3}^{\gamma_3} = 
    \begin{pmatrix}
        1 & -\frac{1}{4\xi} 
        \\
        0 & -1 
    \end{pmatrix}_{\! \alpha_3}^{\! \gamma_3}
    , \q    
    \big(\F{\act}^{001}_0 \big)_{\alpha_3}^{\gamma_3} = 
    \begin{pmatrix}
        1 & -1 
        \\
        0 & 1 
    \end{pmatrix}_{\! \alpha_3}^{\! \gamma_3} ,
    \\
    \big(\F{\act}^{000}_1 \big)_{\alpha_3}^{\gamma_3} = 
    \begin{pmatrix}
        0 & 1 
        \\
        1 & -1 
    \end{pmatrix}_{\! \alpha_3}^{\! \gamma_3}
    , \q 
    \big(\F{\act}^{001}_1 \big)_{\alpha_3}^{\gamma_3} = 
    \begin{pmatrix}
        0 & -1 
        \\
        1 & -\frac{1}{4\xi} 
    \end{pmatrix}_{\! \alpha_3}^{\! \gamma_3} ,
    \\ 
    \big(\F{\act}^{110}_0 \big)_{\alpha_3}^{\gamma_3} = 
    \begin{pmatrix}
        0 & 1 
        \\
        1 & -\frac{1}{4\xi}
    \end{pmatrix}_{\! \alpha_3}^{\! \gamma_3} 
    , \q    
    \big(\F{\act}^{111}_0 \big)_{\alpha_3}^{\gamma_3} = 
    \begin{pmatrix}
        0 & 1 
        \\
        -1 & 1 
    \end{pmatrix}_{\! \alpha_3}^{\! \gamma_3} ,
    \\
    \big(\F{\act}^{110}_1 \big)_{\alpha_3}^{\gamma_3} = 
    \begin{pmatrix}
        -1 & 1 
        \\
        0 & 1 
    \end{pmatrix}_{\! \alpha_3}^{\! \gamma_3}
    , \q 
    \big(\F{\act}^{111}_1 \big)_{\alpha_3}^{\gamma_3} = 
    \begin{pmatrix}
        1 & -\frac{1}{4\xi} 
        \\
        0 & 1
    \end{pmatrix}_{\! \alpha_3}^{\! \gamma_3} .
\end{split}
\end{equation*}
Crucially, these symbols are not unique. Indeed, performing the gauge transformations
\begin{align}
    \raisebox{-1pt}{\actionT{\phi^{\alpha}}{}{\gamma_2}{\gamma_1}} \!\!\!\! \mapsto \, U^{\alpha \gamma_1}_{\gamma_2} \! \raisebox{-1pt}{\actionT{\phi^{\alpha}}{}{\gamma_2}{\gamma_1}} \!\!\! ,
\end{align}
where $U^{\alpha \gamma_1}_{\gamma_2} \in \mathbb{C}^\times$, for every $\gamma_1, \gamma_2, \alpha \in \{0,1\}$, leaves eq.~\eqref{eq:assocAction}
invariant. But, these gauge transformations modify in particular the $\F{\act}$-symbols in the following way:
\begin{equation}
    \big(\F{\act}^{\alpha_1\alpha_1\gamma_1}_{\gamma_2} \big)_{\alpha_2}^{\gamma_3} \mapsto
    \big(\F{\act}^{\alpha_1\alpha_1\gamma_1}_{\gamma_2} \big)_{\alpha_2}^{\gamma_3} 
    \, U^{\alpha_1 \gamma_1}_{\gamma_3} 
    \, U^{\alpha_1 \gamma_3}_{\gamma_2}
    \, \bar U^{\alpha_2 \gamma_1}_{\gamma_2} .
\end{equation}
Equivalence classes of $\F{\act}$-symbols related by gauge transformations classify the different ways ground states transform under the $\Mod(\mc T_4)$ symmetry. However, $\F{\act}$-symbols associated with distinct values of $\xi$ fall within distinct equivalence classes. 
Indeed, it is sufficient to show that we cannot modify $\xi$ to $\xi'$ by gauge transformations without changing any of the other $\F{\act}$-symbols.
Consider the following four entries:
\begin{equation}
    \big(\F{\act}^{000}_0 \big)^{0}_0 =
    \big(\F{\act}^{001}_1 \big)^{1}_0 =
    \big(\F{\act}^{110}_0 \big)^{0}_1 = 
    \big(\F{\act}^{111}_1 \big)_1^{1} = 1 \, .
\end{equation}
In order for these entries to remain equal to 1, we must have
\begin{equation}
U^{0 0}_0 = U^{0 1}_1 = U^{1 0}_0 = U^{1 1}_1 = 1 \, .
\end{equation}
From this, we can already conclude that the value of $\xi$ in the symbols $\big(\F{\act}^{000}_0 \big)^{0}_1$ and $\big(\F{\act}^{001}_1 \big)^{1}_1$ cannot be modified without altering other entries that do not depend on $\xi$. This completes the argument.
In app.~\ref{app:moduleCat}, we relate this statement to the mathematical fact that the non-semisimple tensor category $\Mod(\mc T_4)$ admits an $\mathbb S^1$-parametrised family of rank 2 semisimple module categories, which are inequivalent as module categories for distinct values of the parameter. In the case of a fusion category symmetry, this would be the indication that Hamiltonians $\widetilde{\mathbb H}(\xi)_\Lambda$ represent distinct $\Mod(\mc T_4)$-symmetric gapped phases \cite{Garre-Rubio:2022uum}. However, these are part of the same smooth path of gapped symmetric Hamiltonians, which for self-adjoint Hamiltonians would be taken as the definition that they belong to the same gapped phase. This tension, which might be traced back to the loss of Hermiticity and the necessity to work with open boundary conditions for the symmetry to be faithful, questions either the definition of a gapped phase in the presence of a non-semisimple symmetry or the classification scheme in terms of indecomposable module categories.

\bigskip \noindent
{\bf Spontaneous breaking of the non-semisimple symmetry:} Let us now study here a one-dimensional quantum lattice model whose non-semisimple $\Mod(\mc T_4)$ symmetry is spontaneously broken down to $\mathbb Z/2\mathbb Z$.

In the previous section, we considered the family of (non-hermitian) Hamiltonians parametrised by $\xi \in \rU(1)$ defined in terms of local operators \eqref{eq:localOpXM}. It turns out that our analysis holds more generally for any $\xi \in \mathbb C^\times$, but we restricted to $\xi \in \rU(1)$ to preserve as much unitarity as possible. We now would like to consider the Hamiltonian $\widetilde{\mathbb H}(0)_\Lambda = - \sum_{\msf i \in \Lambda} \widetilde{\mathbb h}(0)_{\msf i,\msf i+1}$ obtained by taking the limit $\xi \to 0$, which is defined in terms of local operators  
\begin{equation}
    \label{eq:localOpW}
    \widetilde{\mathbb h}(0)_{\msf i, \msf i+1} := \mathbb I_\msf i \otimes (\sigma^+ \sigma^-)_{\msf i+1} + \sigma^+_\msf i \otimes \sigma^-_{\msf i+1}  \, .
\end{equation}
The Hamiltonian $\widetilde{\mathbb H}(0)_\Lambda$ retains much of the features of its $\widetilde{\mathbb H}(\xi)_\Lambda$ counterparts: It is non-hermitian and the local operators do not commute with one another. Yet, it is frustration free, presents a spectral gap, and its eigenvalues can be verified to be all real negative. Moreover, it possesses the same non-invertible non-semisimple $\Mod(\mc T_4)$ symmetry. However, the ground state subspace widely differs. 

The ground state subspace of the Hamiltonian $\widetilde{\mathbb H}(0)_\Lambda$ is spanned by two states, namely the product state $|0\ra^{\otimes |\Lambda|}$ and the matrix product state
\begin{equation}
    \label{eq:MPSW}
    \la 1 | \! \raisebox{1.7pt}{\MPS{\rho}} \! | 0 \ra
\end{equation}
where
\begin{align}
    \MPST{\rho}{}{}{}{}{} \!\!\! \equiv \!\!\!
    \sum_{\substack{b \in \{0,1\} \\ c_1,c_2 \in \{0,1\}}} \hspace{-1.2em} \MPST{\rho}{b}{}{}{\, c_1}{c_2 \,}  |b \ra \otimes |c_2 \ra \la c_1 | 
\end{align}
is such that (see sec.~\ref{app:sym_States} for motivation)
\begin{align*}
    \MPST{\rho}{0}{}{}{}{} \!\!\!\! :=  \mathbb I
    \, , \q \!\!
    \MPST{\rho}{1}{}{}{}{} \!\!\!\! := \sigma^- \, .
\end{align*}
We recognise eq.~\eqref{eq:MPSW} as the so-called W state \cite{PhysRevA.62.062314}:
\begin{equation}
    |\text{W} \ra_\Lambda := \sum_{\msf i \in \Lambda} \sigma^-_\msf i \, | 0 \ra^{\otimes |\Lambda|} \, ,
\end{equation}
which one can explicitly check to be a ground state of $\widetilde{\mathbb H}(0)_\Lambda$.\footnote{At this point, it is interesting to note the resemblance between our Hamiltonian and the ferromagnetic XX model with strong magnetic transverse field, which is also a parent Hamiltonian for $|0\ra^{\otimes |\Lambda|}$ and $|\text{W}\ra_\Lambda$, but it is gapless \cite{Perez-Garcia:2006nqo}.}
It follows from $(\widehat \omega_0)_0^1 \, | 0 \ra^{\otimes |\Lambda|} = |\text{W}\ra_\Lambda$ that the symmetry $\Mod( \mc T_4)$ is spontaneously broken down to $\mathbb Z / 2 \mathbb Z$ in the ground state subspace.

It is interesting to revisit certain properties of the W state from the viewpoint of this symmetry breaking pattern. Firstly, the W state cannot be parametrised as a translation invariant matrix product state with tensors of constant size for periodic boundary conditions \cite{Perez-Garcia:2006nqo,klimov2023}. This is a fact that echoes the need to work on open boundary conditions for the $\Mod(\mc T_4)$ symmetry to be well-defined. Secondly, it was recently shown in ref.~\cite{Gioia:2023adm} that the W state cannot be the single ground state of a local Hamiltonian, and must always be accompanied by $|0\ra^{\otimes |\Lambda|}$. In sec.~\ref{app:moduleCat}, we relate this statement to the mathematical fact that $\Mod(\mc T_4)$ admits a non-semisimple module category with two indecomposable objects: a \emph{simple} object and its \emph{projective cover}, labelling the product state and the W state, respectively. 
In contrast to the semisimple setting, there exists non-zero maps between the simple object and the projective object.
In particular, any module category containing the projective object will also have to contain the simple object, which appears as a quotient (or a sub)object.\footnote{Due to this identification, we conjecture that issues arising when dealing with periodic boundary conditions are related to the necessity to define \emph{modified traces} when constructing topological invariants  from non-semisimple tensor categories \cite{geer2009modified,geer2011generalized,orozco2018module,DeRenzi:2019iwu,DeRenzi:2020nzl}. Without these modified traces, the quantum dimension of projective objects would be zero.}
Thirdly, although the W state is long-range entangled, it is indistinguishable from the product state $|0\ra^{\otimes |\Lambda|}$ in the infinite volume limit \cite{Naaijkens:2013vqa}. 
This fact is made possible by the aforementioned existence of maps between the simple and projective objects in the relevant non-semisimple module category, providing a topological local operator, namely $\sigma_\msf i^+$, for any $\msf i \in \Lambda$, mapping the W state to the product state. As such, both ground states should correspond to the same infrared \emph{vacuum} (see sec.~\ref{app:sym_States}).
Therefore, we could argue that this module category does not label a gapped phase distinct from the trivially symmetric one, which would be consistent with the fact that the degeneracy should not be robust to perturbations \cite{Gioia:2023adm}.

\bigskip \noindent
{\bf Discussion:} In this manuscript, we set out to explore through a simple example some consequences of dropping the semisimplicity requirement in the axiomatisation of finite symmetries in (1+1)d in terms of fusion categories. First of all, we noticed that such a non-semisimple symmetry seems to be incompatible with Hermiticity of the Hamiltonian. This fact was to be anticipated in light of previous instances of similar phenomena \cite{Geer:2022qoa}, and further requires open boundary conditions. Nonetheless, this did not prevent us from finding certain frustration free Hamiltonian operators with real spectra. 

We examined two scenarios that refines the current paradigm for the classification of gapped symmetric phases in terms of indecomposable module categories over the symmetry category. On the one hand, we obtained a continuous family of product states that transform inequivalently under the non-semisimple symmetry, a phenomenon that cannot occur in the case of a finite semisimple symmetry. This is explained mathematically by a continuum of inequivalent semisimple (exact) module categories. On the other hand, we found a gapped model with two symmetry breaking ground states, which happen to be indistinguishable in the infinite volume limit. We trace this phenomenon back to one of the ground states being associated with an indecomposable object that is the \emph{projective cover} of  the other. 
\emph{Dicke states} \cite{PhysRev.93.99} that generalise the W state also seem to be related to higher order Taft algebras in a similar vein. 
In fact, we expect this to be a common phenomenon in non-semisimple module categories.
As such, it would also be interesting to explore the physical interpretation of non-semisimple module categories even in the context of fusion categories \cite{EKW:2021ta,Heng:2024cox,EliasHeng:2024fq}.

In the appendices, we explore the physical content of additional indecomposable module categories over $\Mod(\mc T_4)$. Notably, a continuous family of fiber functors produces an $\mathbb S_1$-parametrised family of states, which are in the same phase as the so-called \emph{cluster state} with respect to a $\mathbb Z / 2 \mathbb Z \times \mathbb Z / 2 \mathbb Z$ symmetry, and yet transform inequivalently
with respect to $\Mod(\mc T_4)$. However, for this family of states, we were unable to find $\Mod(\mc T_4)$-symmetric parent Hamiltonians with real spectra.

\bigskip\noindent
\textbf{Acknowledgments:} \emph{CD is grateful to Laurens Lootens and Alex Turzillo for useful discussions. EH is grateful for the collaborative opportunity made possible through the Institut des Hautes \'{E}tudes Scientifiques (IHES). MY is supported by the EPSRC Open Fellowship EP/X01276X/1, and would like to thank Thibault Décoppet and Thomas Wasserman for helpful discussions as well as the IHES for hosting visits where part of this research was conducted.
}

\titleformat{name=\section}
{\normalfont}
{\centering {\textsc{App.} \thesection \; \raisebox{1pt}{\textbar} \;}}
{0pt}
{\normalsize\bfseries\centering}

\newpage
\appendix
\section{Category theoretic underpinnings}

\noindent
\emph{We present in these appendices the mathematical formalism underlying our study, and exploit this formalism to further elucidate the results enunciated in the main text. Even though we specialise to the case of the Taft algebra, most of the constructions presented in these appendices hold much more generally.}

\subsection{Taft algebra $\mc T_4$\label{app:Taft}}

\noindent
Let $\mc T_4$ be the \emph{Taft Hopf algebra} of dimension 4 \cite{Taft}, also known as Sweedler's Hopf algebra \cite{sweedler1969hopf}, which is the lowest dimensional Hopf algebra that is both non-commutative and non-cocommutative.\footnote{For a brief review of Hopf algebra theory, see e.g. \cite{Montgomery,DNR,CK}.} As an associative algebra, it is
\begin{equation}
    \label{eq:H2alg}
    \mc T_4 = \mathbb C \la x,g \, | \, x^2=0, \, g^2=1, \, xg=-gx \ra\, .
\end{equation}
The comultiplication $\Delta: \mc T_4 \to \mc T_4 \otimes \mc T_4$ and counit $\epsilon: \mc T_4 \to \mathbb C$ are given by 
\begin{equation}
\begin{split}
    \label{eq:H2coalg}
    \Delta(g) &= g \otimes g \, , \q \Delta(x) = x \otimes 1 + g \otimes x  \, ,
    \\
    \epsilon(g) &= 1 \, , \q\q\q\, \epsilon(x) =0 \, ,
\end{split}
\end{equation}
respectively, which provide the coalgebraic structure. Finally, the antipode $S: \mc T_4 \to \mc T_4$ defined by
\begin{equation}
    \label{eq:H2Hopf}
    S(g) = g \, , \q S(x) =xg 
\end{equation}
endows the resulting bialgebra with its Hopf algebraic structure. Notice, in particular, that $S^2 \neq \text{id}$. Loosely, we can think of $\mc T_4$ as a minimal non-semisimple extension of the group algebra $\mathbb C[\mathbb Z / 2 \mathbb Z]$ by $\mathbb C[x]/\{x^2=0\}$. Throughout these appendices, we employ Sweedler's sumless notation for coalgebraic structures, e.g. $\Delta(x) \equiv x_{(1)} \otimes x_{(2)}$.

\subsection{Tensor category $\Mod(\mc T_4)$\label{app:TC}}

\noindent
Let us construct the tensor category $\Mod(\mc T_4)$ of left modules over the Taft algebra $\mc T_4$. Given the associative algebra structure \eqref{eq:H2alg}, let us begin by listing the \emph{indecomposable} modules over $\mc T_4$.  We emphasise here that an indecomposable object need not be simple; this is in contrast to the semisimple setting, where an object is simple if and only if it is indecomposable. In particular, two distinct indecomposable objects in $\Mod(\mc T_4)$ may have non-zero maps between them.

Firstly, there are two \emph{simple} one-dimensional modules: 
\begin{align*}
    S_0 = \mathbb C\{w_1\} \q \text{w/} \;
    \left\{
    \begin{array}{l}
        x \cdot w_1 = 0  
        \\
        g \cdot w_1 = w_1  
    \end{array}
    \right. \! ,
\end{align*}
which plays the role of the \emph{trivial} module, and
\begin{align*}
    S_1 = \mathbb C\{v_1\} \q \text{w/} \; 
    \left\{
    \begin{array}{l}
        x \cdot v_1 = 0  
        \\
        g \cdot v_1 = -v_1  
    \end{array}
    \right. \! .
\end{align*}
Secondly, there are two \emph{projective} modules:
\begin{align*}
    P_0 = \mathbb C\{v_0,v_1\} \q \text{w/} \;  
    \left\{
    \begin{array}{ll}
        x \cdot v_0 = v_1   , \q & x \cdot v_1 =0
        \\
        g \cdot v_0 = v_0  , & g \cdot v_1 = -v_1 
    \end{array}
    \right.
\end{align*}
and
\begin{align*}
    P_1 = \mathbb C\{w_0, w_1\} \q \text{w/} \; 
    \left\{
    \begin{array}{ll}
        x \cdot w_0 = w_1 \, , \q & x \cdot w_1 = 0  
        \\
        g \cdot w_0 = -w_0 \, , & g \cdot w_1 = w_1  
    \end{array}
    \right. \!\! .
\end{align*}
Being indecomposable projective modules, $P_0$ and $P_1$ cannot be written as direct sums of simple objects.
Instead, each $P_0$ and $P_1$ are the \emph{projective covers} of $S_0$ and $S_1$, respectively, and they fit into the short exact sequences
\begin{equation}
\begin{split}
    \label{eq:H2seq}
    0 &\to S_1 \to P_0 \to S_0 \to 0 \, , 
    \\
   0 &\to S_0 \to P_1 \to S_1 \to 0 \, .
\end{split}
\end{equation}
Note that there are non-zero maps from $P_0$ to $P_1$ and $P_1$ to $P_0$ induced by factoring through the quotient maps onto $S_0$ and $S_1$, respectively.
Moreover, $\mc T_4$ is isomorphic to $P_0 \oplus P_1$ as objects in $\Mod(\mc T_4)$.

The coalgebraic structure provided in eq.~\eqref{eq:H2coalg} yields the monoidal structures given by
\begin{equation*}
\begin{split}
    S_1 \otimes S_1 &\cong S_0 
    \cong \mathbb C\{v_1 \otimes v_1\} \, , 
    \\
    P_0 \otimes S_1 &\cong P_1
    \cong \mathbb C\{v_0 \otimes v_1 , -v_1 \otimes v_1\} \, ,
    \\
    P_1 \otimes S_1 &\cong P_0 
    \cong \mathbb C\{w_0 \otimes v_1, w_1 \otimes v_1\} \, ,
    \\
    S_1 \otimes P_0 &\cong P_1 
    \cong \mathbb C\{v_1 \otimes v_0, -v_1 \otimes v_1\} \, ,
    \\
    S_1 \otimes P_1 &\cong P_0
    \cong \mathbb C\{v_1 \otimes w_0 , v_1 \otimes w_1\} \, ,
\end{split}
\end{equation*}
and 
\begin{align} \label{eqn:H4projectivedecomp}
    \nn
    P_0 \otimes P_0 \cong P_0 \oplus P_1 
    \cong \; &\mathbb C\{v_0 \otimes v_0, v_0 \otimes v_1 + v_1 \otimes v_0\} 
    \\ 
    &\! \oplus \mathbb C\{v_1 \otimes v_0, -v_1 \otimes v_1\} \, ,
    \\ \nn 
    P_0 \otimes P_1 \cong P_0 \oplus P_1 
    \cong \; &\mathbb C\{v_1 \otimes w_0, -v_1 \otimes w_1\} 
    \\ \nn
    &\! \oplus \mathbb C\{v_0 \otimes w_0, v_0 \otimes w_1 + v_1 \otimes w_0\} \, ,
    \\ \nn
    P_1 \otimes P_0 \cong P_0 \oplus P_1 
    \cong \; &\mathbb C\{w_0 \otimes v_1, w_1 \otimes v_1\} 
    \\ \nn 
    &\! \oplus \mathbb C\{w_0 \otimes v_0, w_1 \otimes v_0 -w_0 \otimes v_1\} \, ,
    \\ \nn
    P_1 \otimes P_1 \cong P_0 \oplus P_1 
    \cong \; &\mathbb C\{w_0 \otimes w_0, w_1 \otimes w_0 - w_0 \otimes w_1\} 
    \\ \nn
    &\! \oplus \mathbb C\{w_1 \otimes w_0, w_1 \otimes w_1\} \, .
\end{align}
Let us emphasise that the identifications above are given by their decompositions into indecomposable objects, which are unique up to isomorphisms; this is known as the \emph{Krull--Schmidt property}, which is satisfied by all finite abelian categories. Such a decomposition agrees with the decomposition into simples in the semisimple setting, where an object is indecomposable if and only if it is simple.

The role of the monoidal unit is played by $S_0$, which is simple. The Hopf algebraic structure provided in eq.~\eqref{eq:H2Hopf} finally yields the rigidity structure
\begin{equation}
    S_1^\vee \cong S_1 \, , \q P_0^\vee 
    \cong P_1 \, , \q P_1^\vee \cong P_0 \, .
\end{equation}
This completes the definition of $\Mod(\mc T_4)$ as a finite tensor category in the sense of ref.~\cite{etingof2003}, which is non-semisimple by virtue of eq.~\eqref{eq:H2seq}.

From here on, we use the notation $V_3 \in V_1 \otimes V_2$ to mean $V_3$ appears as a summand in the tensor product $V_1 \otimes V_2$.
From the monoidal structures computed above, we obtain intertwining maps $\varphi^{V_1  \! V_2}_{V_3} \in \Hom_{\Mod(\mc T_4)}(V_1 \otimes V_2,V_3)$ for each indecomposable $V_3\in V_1 \otimes V_2$. 
Choosing bases $V_1 = \mathbb C\{v_{d_1}\}_{d_1}$, $V_2 = \mathbb C\{v_{d_2}\}_{d_2}$ and $V_3 = \mathbb C\{v_{d_3}\}_{d_3}$, components of the linear map $\varphi^{V_1 \! V_2}_{V_3}$ are denoted by $\big(\varphi^{V_1 \! V_2}_{V_3} \big)_{d_1 d_2}^{d_3} \in \mathbb C$. This allows us to define the following `fusion' tensors:
\begin{align}
    \nn
    \fusionT{\;\;\;\, \varphi^{V_1 \! V_2}_{V_3}}{}{}{} 
    \!\! &\equiv \! 
    \sum_{d_1,d_2,d_3} \!\! \fusionT{\;\;\;\, \varphi^{V_1 \! V_2}_{V_3}}{\, d_1}{\, d_2}{d_3} \, v_{d_3} \otimes v_{d_1}^* \otimes v_{d_2}^* 
    \\
    &\equiv 
    \! \sum_{d_1,d_2,d_3} \big(\varphi^{V_1 \! V_2}_{V_3} \big)_{d_1 d_2}^{d_3} \, v_{d_3} \otimes v_{d_1}^* \otimes v_{d_2}^* \, .
\end{align}
Moreover, the monoidal structures above also provide us with linear maps $\bar \varphi^{V_1 \! V_2}_{V_3} : V_3 \to V_1 \otimes V_2$ satisfying orthogonality conditions
\begin{equation}
    \orthoFusionGen = \delta_{V_3,V_4} \mathbb I_{V_3} \, .
\end{equation}
In the main text, tensors \eqref{eq:fusion_P0} and \eqref{eq:fusion_P1} precisely corresponds to the linear maps
$\varphi^{00}_\alpha \equiv \varphi^{P_0 P_0}_{P_\alpha} : P_0 \otimes P_0 \to P_\alpha$ and $\varphi^{11}_\alpha \equiv \varphi^{P_1 P_1}_{P_\alpha} : P_1 \otimes P_1 \to P_\alpha$, respectively, under the identifications $v_{d_1} \equiv |d_1 \ra$ and $w_{d_2} \equiv |d_2 \ra$, for every $d_1,d_2 \in \{0,1\}$.

\subsection{Spin chains with $\Mod(\mc T_4)$ symmetry\label{app:chain}}

\noindent
Let us now explain how to construct one-dimensional quantum lattice models hosting a $\Mod(\mc T_4)$ symmetry. In ref.~\cite{Lootens:2021tet,Lootens:2022avn}, it was established that the action of a finite tensor category $\mc C$ on a discrete quantum mechanical system is specified by the data of a (right) indecomposable \emph{module category} $\mc M$ and of an object in the \emph{Morita dual} $\mc C^\vee_\mc M$ of $\mc C$ with respect to $\mc M$, which is defined as the category of module endofunctors of $\mc M$ over $\mc C$.\footnote{See ref.~\cite{etingof2016tensor} for an introduction to module category theory.} Whenever the symmetry is \emph{non-anomalous} so that the tensor product admits a \emph{fiber functor}---i.e., a module category over $\mc C$ that is equivalent to the category $\Vect$ of complex vector spaces---it is possible to realise the action of $\mc C$ on a tensor product Hilbert space. Choosing the fiber functor to be the \emph{forgetful functor} $\msf{Forg} : \Mod(\mc T_4) \to \Vect$ recovers the framework of ref.~\cite{Inamura:2021szw}. We always make this choice in what follows.\footnote{Choosing different fiber functors would ultimately lead to quantum lattice models that are dual to one another according to ref.~\cite{Lootens:2021tet,Lootens:2022avn}.}

Since the Morita dual of $\Mod(\mc T_4)$ with respect to $\Vect$ is equivalent to the (finite tensor) category $\Comod(\mc T_4)$ of (left) comodules over $\mc T_4$, we must also make a choice of $\mc T_4$-\emph{comodule}. By definition, a left $\mc T_4$-comodule is a vector space $K$ together with a left \emph{coaction} $\lambda : K \to \mc T_4 \otimes K$ that is \emph{coassociative}, i.e.
\begin{equation}
    (\Delta \otimes {\rm id}_{K}) \circ \lambda = ({\rm id}_{\mc T_4} \otimes \lambda) \circ \lambda \, ,
\end{equation}
and \emph{unital}
\begin{equation}
    (\epsilon \otimes {\rm id}_K ) \circ \lambda = {\rm id}_K \, .    
\end{equation}
Using Sweedler's sumless notation, one writes $\lambda(k) = k_{(-1)} \otimes k_{(0)}$, for every $k \in K$. More explicitly, given a choice of basis $\{k_{b}\}_b$, $b=0,\ldots,\dim_\mathbb C K-1$, one writes $\lambda(k_{b_1}) = \sum_{b_2} \lambda_{b_1}^{b_2} \otimes k_{b_2}$, where $\lambda_{b_1}^{b_2} \in \mc T_4$ is not necessarily a basis vector. In this notation, the coassociativity of the coaction translates into
\begin{equation}
    \label{eq:coassoc_coaction}
    \Delta(\lambda_{b_1}^{b_3}) = \sum_{b_2} \lambda_{b_1}^{b_2} \otimes \lambda_{b_2}^{b_3}\,.
\end{equation}
Given a finite subset $\Lambda$ of the lattice $\mathbb Z$, one defines the microscopic Hilbert space of the system to be $\mc H_\Lambda = \bigotimes_{\msf i \in \Lambda} \msf{Forg}(K)$. From ref.~\cite{Lootens:2021tet,Lootens:2022avn}, we know that the action of $\Mod(\mc T_4)$ on this Hilbert space is provided by the data of module endofunctors in ${(\Comod(\mc T_4))}^\vee_\Vect \simeq \Mod(\mc T_4)$. Concretely, given $(\varrho: \mc T_4 \to \End(V)) \in \Mod(\mc T_4)$, let
\begin{equation}
    \arraycolsep=1.4pt
    \begin{array}{ccccl}
        \omega & : & K \otimes V & \to & V \otimes K
        \\
        & & k \otimes v & \mapsto & (k_{(-1)} \cdot v) \otimes k_{(0)} 
    \end{array} \, . 
\end{equation}
In components, it reads
\begin{equation*}
    \label{eq:omegabasis}
    k_{b_1} \otimes v_{d_1} 
    \mapsto \sum_{b_2} (\lambda_{b_1}^{b_2} \cdot v_{d_1}) \otimes k_{b_2}
    = \sum_{b_2,d_2} \varrho(\lambda_{b_1}^{b_2})_{d_1}^{d_2} \; v_{d_2} \otimes k_{b_2} \, ,
\end{equation*}
where $\{v_d\}_d$, $d=0,\ldots, \dim_\mathbb C V-1$, is a choice of basis for $V$. 
Introducing the notation\footnote{Notice that we are transposing some indices for the need of our presentation.}
\begin{equation}
    \label{eq:MPOT}
    \omega^{d_2}_{d_1} := \sum_{b_1,b_2} \varrho(\lambda_{b_2}^{b_1})_{d_1}^{d_2} \; k_{b_2} \otimes k_{b_1}^* 
    \, ,
\end{equation}    
the action of $\varrho : \mc T_4 \to \End(V)$ in $\Mod(\mc T_4)$ on $\mc H_\Lambda$ with open boundary conditions $v_{d_1},v_{d_{|\Lambda|+1}} \in V$ is defined to be
\begin{equation}
    \sum_{d_2,\ldots,d_{|\Lambda|}} \!\!\!
    \omega^{d_{|\Lambda|+1}}_{d_{|\Lambda|}} \otimes \cdots \otimes \omega^{d_3}_{d_2} \otimes \omega^{d_2}_{d_1} \, . 
\end{equation}
This operator can be expressed as a tensor network of the form \eqref{eq:MPO} by defining the following tensor:
\begin{align}
    \nn
    \MPOT{\! \omega}{}{}{}{} \!\!\!\!
    &\equiv \sum_{\substack{d_1,d_2 \\ b_1,b_2}} \!
    \MPOT{\! \omega}{b_2}{\, d_1}{d_2 \,}{b_1} \, 
    k_{b_2}  \otimes k_{b_1}^* \otimes v_{d_2} \otimes v^*_{d_1}
    \\
    &\equiv
    \sum_{\substack{d_1,d_2 \\ b_1,b_2}} 
     \varrho(\lambda_{b_2}^{b_1})_{d_1}^{d_2} \; 
    k_{b_2}  \otimes k_{b_1}^* \otimes v_{d_2} \otimes v^*_{d_1} \, .
\end{align}
Let us consider an explicit example. Let $K = \mathbb C\{1,y\}$ with $\lambda : K \to \mc T_4 \otimes K$ such that $\lambda(y) = x \otimes 1 + g \otimes y$. One can explicitly check that $\lambda$ is coassociative and counital so that $K$ has the structure of a left $\mc T_4$-comodule. Now choose the object in $\Mod(\mc T_4)$ to be the projective object $P_0$ defined previously. Applying the definitions, one finds
\begin{equation}
\begin{split}
    \omega : y \otimes v_0 &\mapsto v_1 \otimes 1 + v_0 \otimes y
    \\
    y \otimes v_1 &\mapsto - v_1 \otimes y 
\end{split} \, .
\end{equation}
Making the identifications $v_d \equiv |d \ra$ and $k_b \equiv |b \ra$, for every $b,d \in \{0,1\}$, one exactly recovers the symmetry operators \eqref{eq:MPO_P0} defined in the main text. Similarly, the simple object $S_1$ yields the symmetry operator $\prod_{\msf i \in \Lambda} \sigma^z_\msf i$, while $P_1$ produces the composition of the symmetry operators associated with $P_0$ and $S_1$, respectively, as predicted by the monoidal structure of $\Mod(\mc T_4)$. Interestingly, the fact that for three choices of open boundary conditions, the symmetry operator labelled by $P_0$ boils down to the identity operator, $\prod_{\msf i \in \Lambda} \sigma_\msf i^z$, and the zero operator, respectively, is explained by the fact that $P_0$ fits into the short exact sequence \eqref{eq:H2seq}.

Let $\varrho_{V_1}: \mc T_4 \to \End(V_1)$ and $\varrho_{V_2} : \mc T_4 \to \End(V_2)$ be two indecomposable objects in $\Mod(\mc T_4)$. By the definition of the coalgebraic structure of $\mc T_4$ and eq.~\eqref{eq:coassoc_coaction}, the intertwining map $\varphi^{V_1 \! V_2}_{V_3} : V_1 \otimes V_2 \to V_3$, for any indecomposable $V_3 \in V_1 \otimes V_2$, satisfies
\begin{equation}
\begin{split}
    &\sum_{b_2,d_2,d_4} 
    \big( \varphi^{V_1 \! V_2}_{V_3}\big)_{d_2d_4}^{d_5} \,
    \varrho_{V_1}(\lambda_{b_1}^{b_2})_{d_1}^{d_2} \, \varrho_{V_2}(\lambda_{b_2}^{b_3})_{d_3}^{d_4} 
    \\
    & \q = \sum_{d_2,d_4} 
    \big( \varphi^{V_1 \! V_2}_{V_3}\big)_{d_2d_4}^{d_5} \,
    (\varrho_{V_1} \otimes \varrho_{V_2})\big(\Delta(\lambda_{b_1}^{b_3})\big)_{d_1 d_3}^{d_2 d_4}  
    \\
    & \q = \sum_{d_6} \,  \varrho_{V_3}(\lambda_{b_1}^{b_3})_{d_6}^{d_5} \, \big( \varphi^{V_1 \! V_2}_{V_3}\big)_{d_1 d_3}^{d_6} \, ,
\end{split}
\end{equation}
from which the following tensor network identity follows:
\begin{equation}
    \fusionMPO{1}{\omega_{V_1}}{\omega_{V_2}}{} =  \!\! \sum_{V_3 \in V_1 \otimes V_2} \! \fusionMPO{2}{\omega_{V_3}}{\! \varphi^{V_1 \! V_2}_{V_3} }{\bar \varphi^{V_1\! V_2}_{V_3}} \! .
\end{equation}
Specialising to $K=\mathbb C\{1,y\}$ as defined above and choosing $V_1 = V_2 = P_0$, $V_3$ can either be $P_0$ or $P_1$. In light of the previous identifications, the above equality then boils down to eq.~\eqref{eq:fusion_P0}. 

Now that we have explained how $\Mod(\mc T_4)$ acts on the microscopic Hilbert space $\mc H_\Lambda = \bigotimes_{\msf i \in \Lambda} \msf{Forg}(K)$, where $K$ is any object in $\Comod(\mc T_4)$, we are left to construct linear operators $\mc H_\Lambda \to \mc H_\Lambda$ that commute with this action. Let $\psi : K^{\otimes 2} \to K^{\otimes 2}$ be a morphism in $\Comod(\mc T_4)$. By definition, $\msf{Forg}(\psi)$ acts on $\msf{Forg}(K)^{\otimes 2}$. Assigning the two copies of $\msf{Forg}(K)$ to sites $\msf i$ and $\msf i+1$ in $\Lambda$, respectively, we denote by $\mathbb h_{\msf i,\msf i+1}$ the embedding of $\msf{Forg}(\psi)$ into $\mc H_\Lambda$. One finally construct a Hamiltonian operator
$\mathbb H_\Lambda = - \sum_{\msf i \in \Lambda} \mathbb h_{\msf i,\msf i+1}$. The fact that $\mathbb H_\Lambda$ commutes with the symmetry follows from the fact that, as a map in $\Comod(\mc T_4)$, $\psi$ commutes with the coaction of $\mc T_4$. In app.~\ref{app:parent}, we explain how to obtain the Hamiltonians considered in the main text within this framework.

\subsection{Algebra objects in $\Comod(\mc T_4)$\label{app:algObj}}

\noindent
Given a symmetry fusion category, it is understood that symmetric gapped phases are labelled by indecomposable (finite semisimple) module categories over it \cite{Thorngren:2019iar,Komargodski:2020mxz}. In that spirit, we are interested in the classification of \emph{exact} indecomposable module categories over $\Mod(\mc T_4)$. In ref.~\cite{Mombelli2010}, it was demonstrated how these could be constructed from the data of so-called \emph{right $\mc T_4$-simple left $\mc T_4$-comodule algebras}.

By definition, a left $\mc T_4$-comodule algebra $\mc A$ is an algebra in $\Vect$, together with a left $\mc T_4$-comodule structure, such that the coaction $\lambda$ is an algebra homomorphism. In other words, these are \emph{algebra objects} in $\Comod(\mc T_4)$ \cite{etingof2003}. Moreover, a left $\mc T_4$-comodule algebra $K$ is said to be \emph{right $\mc T_4$-simple} whenever the only right $\mc T_4$-ideal $J$ of $K$ with the property that $\lambda(J) \subset \mc T_4 \otimes K$, is the trivial ideal. We review below the classification of the resulting right $\mc T_4$-simple left $\mc T_4$-comodule algebras as established in ref.~\cite{Mombelli2010}.

We distinguish two types of right $\mc T_4$-simple left $\mc T_4$-comodule algebras. On the one hand, we have the group algebras $\mathbb C[\mathbb Z/1\mathbb Z] \cong \mathbb C$ and $\mathbb C[\mathbb Z/2 \mathbb Z] = \mathbb C\la h \, | \, h^2 = 1\, \ra$, which are semisimple. On the other hand, for any $\xi \in \mathbb C$, one defines
\begin{equation}
    \label{eq:AdXi}
    \begin{split}
        \mc A(1,\xi) &= \mathbb C\la y \, | \, y^2=\xi \cdot 1 \ra \, ,
        \\
        \mc A(2,\xi) &= \mathbb C\la y,h \, | \, y^2=\xi \cdot 1, \, h^2=1, \, yh=-hy\ra \, ,
    \end{split}
\end{equation}
 together with the comodule structure $\lambda$ provided by
\begin{equation}
    \label{eq:coaction}
    \lambda(h) = g \otimes h \, , \q \lambda(y) = x \otimes 1 + g \otimes y \, .
\end{equation}
Let us analyse this latter type is some detail. First of all, $\mc A(1,\xi)$ and $\mc A(2,\xi)$ are non-semisimple if and only if $\xi = 0$. As a matter of fact $\mc A(2,0)$ is the Taft algebra $\mc T_4$ itself, whereas $\mc A(1,0)$ is the left coideal subalgebra of $\mc T_4$ generated by $x$. Let us suppose that $\xi \in \mathbb C^\times$. Clearly $\mc A(2,0)$ and $\mc A(2,\xi)$ are not isomorphic as algebras. However, $\mc A(2,\xi)$ can be realised as a twisted version of $\mc A(2,0)$. 

Let $\beta_\xi : \mc T_4 \otimes \mc T_4 \to \mathbb C$ be the function such that $\beta_\xi(1,g)=\beta_\xi(g,1)=\epsilon(g)$, $\beta_\xi(1,x)=\beta_\xi(x,1)=\epsilon(x)$, $\beta_\xi(x,x)=\xi$ and 
\begin{equation}
    \beta_\xi(x^{a_1}g^{a_2},x^{a_3}g^{a_4}):= (-1)^{a_2a_3}\beta_\xi(x^{a_1},x^{a_3}) \, ,
\end{equation}
for every $a_1,a_2,a_3,a_4 \in \{0,1\}$. One can verify that $\beta_\xi$ satisfies the following condition
\begin{equation*}
    \beta_\xi(a_{(1)}, b_{(1)}) \, \beta_\xi(a_{(2)}b_{(2)}, c) = \beta_\xi(b_{(1)}, c_{(1)})
    \beta_\xi(a ,  b_{(2)}c_{(2)}) \, ,
\end{equation*} 
for every $a,b,c \in \mc T_4$, which is the defining property of a \emph{Hopf 2-cocycle}. Moreover, 
given $\xi,\xi' \in \mathbb C^\times$ such that $\xi \neq \xi'$, one can verify that the Hopf 2-cocycles $\beta_\xi$ and $\beta_{\xi'}$ of $\mc T_4$ are inequivalent, in the sense that there is no \emph{convolution unit} $\theta$ of the dual of $\mc T_4$ such that $\beta_{\xi'}(a,b)$ is equal to
\begin{equation*}
    \theta(a_{(1)}) \, \theta(b_{(1)}) \, \beta_\xi(a_{(2)},b_{(2)}) \, \theta^{-1}(a_{(3)}b_{(3)}) \, ,
\end{equation*}
for every $a,b \in \mc T_4$ \cite{Garcia}. 

One can use the Hopf 2-cocycle $\beta_\xi$ to define a twisted multiplication rule for the algebra $\mc A(2,0)$ via its $\mc T_4$-comodule structure according to
\begin{equation}
    a \cdot_{\beta_\xi} b := \beta_\xi(a_{(-1)},b_{(-1)}) \, a_{(0)} \cdot b_{(0)} \, ,
\end{equation}
for every $a,b\in \mc A(2,0)$. The defining property of $\beta_\xi$ ensures that this multiplication rule is associative. One can easily verify that $\mc A(2,0)$ equipped with the twisted multiplication rules $\cdot_{\beta_\xi}$ is indeed isomorphic to $\mc A(2,\xi)$. Interestingly, even though $\beta_\xi$ and $\beta_{\xi'}$ are inequivalent Hopf 2-cocycles for $\xi \neq \xi'$, the resulting algebras (in $\Vect$) are isomorphic. Similarly, twisting the multiplication rule of $\mc A(1,0)$ by $\beta_\xi$ results in an algebra isomorphic to $\mc A(1,\xi)$.

At this point, it is interesting to note that the functions $\beta_\xi$, for every $\xi \in \mathbb C^\times$, also define group 2-cocycles that can be used to define twisted multiplication rules for the group algebra $\mathbb C[\mathbb Z/  2 \mathbb Z \times \mathbb Z / 2 \mathbb Z] = \mathbb C\la x,g \, | \, x^2=1=g^2 \ra$, resulting in a twisted group algebra that is isomorphic to $\mc A(2,\xi)$. However, for every $\xi \in \mathbb C^\times$, these group 2-cocycles all fall within the same cohomology class.

In the following, we compute the exact indecomposable module categories over $\Mod(\mc T_4)$ associated with the algebra objects in $\Comod(\mc T_4)$ listed above. 

\subsection{Module categories over $\Mod(\mc T_4)$\label{app:moduleCat}}

\noindent
Every exact indecomposable module category $\mc M$ over $\Mod(\mc T_4)$ is equivalent to the category of left modules $\Mod(\mc A)$ over one of the right $\mc T_4$-simple left $\mc T_4$-comodule algebras $\mc A$ constructed above, such that the module structure of $\Mod(\mc A)$ is provided by the $\mc T_4$-comodule structure of $\mc A$ \cite{Mombelli2010}.\footnote{Notice that while the category $\Mod(\mc A)$ of $\mc A$-modules (in $\Vect$) has the structure of a left $\Mod(\mc T_4)$-module category, the category $\Mod_{\Comod(\mc T_4)}(\mc A)$ of $\mc A$-modules in $\Comod(\mc T_4)$ would have the structure of a right $\Comod(\mc T_4)$-module category.} Most importantly, $\Mod(\mc A(n,\xi))$ and $\Mod(\mc A(n',\xi'))$ are equivalent as $\Mod(\mc T_4)$-module categories if and only if $n=n'$ and $\xi = \xi'$, which can be traced back to $\beta_\xi$ and $\beta_{\xi'}$ being inequivalent Hopf 2-cocycles as long as $\xi \neq \xi'$ \cite{Mombelli2010}. Therefore, the algebra objects listed in app.~\ref{app:algObj} classify the exact indecomposable module categories \cite{Mombelli2010,etingof2003}.

More concretely, given a right $\mc T_4$-simple left $\mc T_4$-comodule algebras $\mc A$, the category $\Mod(\mc A)$ can be equipped with an action bifunctor
\begin{equation}
    \act : \Mod(\mc T_4) \times \Mod(\mc A) \to \Mod(\mc A)
\end{equation}
as well as a \emph{module associator} $\F{\act}$ specified by a collection of isomorphisms 
\begin{equation}
    \F{\act}^{V_1V_2M}: (V_1 \otimes V_2) \act M \xrightarrow{\sim} V_1 \act (V_2 \act M) 
\end{equation}
satisfying a \emph{pentagon axiom}, for every $V_1,V_2 \in \Mod(\mc T_4)$ and $M \in \Mod(\mc A)$. Given $V \in \Mod(\mc T_4)$ and $M \in \Mod(\mc A)$, $V \act M$ is simply defined as the tensor product $V \otimes M$ in $\Vect$, and the $\mc A$-module structure on $V \otimes M$ is provided by the coaction $\lambda : \mc A \to \mc T_4 \otimes \mc A$  via $k \cdot (v \otimes m) := (k_{(-1)} \cdot v ) \otimes (k_{(0)} \cdot m)$, for every $k \in \mc A$, $v \in V$ and $m \in M$.
Furthermore, the module associator $\F{\act}$ is simply provided by the associativity in $\Vect$. Below, we work out in some detail the $\Mod(\mc T_4)$-module structures of categories $\Mod(\mc A(1,0))$, $\Mod(\mc A(1,\xi))$ and $\Mod(\mc A(2,\xi))$, for any $\xi \in  \mathbb C^\times$:

\bigskip\noindent
$\bul$ $\Mod(\mc A(1,0))$: Given the associative algebra structure \eqref{eq:AdXi}, this is a rank 2 non-semisimple category whose two indecomposable objects are the one-dimensional simple  module
\begin{equation*}
    S = \mathbb C\{m_1\} \q \text{w/} \q y \cdot m_1 = 0 
\end{equation*}
and the two-dimensional projective module
\begin{equation*}
    P = \mathbb C\{m_0,m_1\} \q \text{w/} \q y \cdot m_0 = m_1 , \q y \cdot m_1 = 0 \, ,
\end{equation*}
respectively. Moreover, $P$ is the projective cover of the unique simple $S$, and it fits into the short exact sequence
\begin{equation}
    0 \to S \to P \to S \to 0 \, .
\end{equation}
Finally, $\Hom_{\Mod(\mc A(1,0))}(P,P) \cong \mc A(1,0)$ with the non-identity (nilpotent) map $P \to P$ factoring through $S$. We will discuss the physical implications of this fact in app.~\ref{app:sym_States}. The $\mc T_4$-comodule structure provided in eq.~\eqref{eq:coaction} then yields the module structure given by
\begin{equation*}
\begin{split}
    S_1 \act S & \cong S \cong \mathbb C\{v_1 \otimes m_1\} \, ,
    \\
    P_0 \act S &\cong P \cong\mathbb{C}\{v_0 \otimes m_1,v_1 \otimes m_1\} \, , 
    \\
    P_1 \act S &\cong P \cong \mathbb{C}\{w_0 \otimes m_1, w_1\otimes m_1\} \, ,
    \\
    S_1 \act P &\cong P \cong \mathbb{C}\{-v_1 \otimes m_0,v_1 \otimes m_1\} \, ,
\end{split}
\end{equation*}
and
\begin{equation*}
\begin{split}
    P_0 \act P \cong P \oplus P \cong \; &\mathbb{C}\{ v_0 \otimes m_0, v_1 \otimes m_0+v_0 \otimes m_1\}
    \\
    &\! \oplus \mathbb{C}\{-v_1 \otimes m_0, v_1 \otimes m_1\} \, ,
    \\
    P_1 \act P \cong P \oplus P \cong \; &\mathbb{C}\{-w_0 \otimes m_0, w_1 \otimes m_0 - w_0 \otimes m_1\}  
    \\
    &\! \oplus \mathbb{C}\{ w_0 \otimes m_1, w_1 \otimes m_1\} \, .
\end{split}
\end{equation*}
As with $\Mod(\mc T_4)$, let us emphasise that the identifications above are given in terms of their decompositions into indecomposable objects (which are unique up to isomorphism).

\bigskip \noindent
$\bul$ $\Mod(\mc A(1,\xi))$: In sharp contrast to the case $\xi=0$, this is a rank 2 semisimple category whose two simple one-dimensional modules are given by
\begin{equation*}
    T_0 = \mathbb C\{m_0\} \q \text{w/} \q y \cdot m_0 = \sqrt{\xi} \, m_0
\end{equation*}
and
\begin{equation*}
    T_1 = \mathbb C\{m_1\} \q \text{w/} \q y \cdot m_1 = -\sqrt{\xi} \, m_1 \, ,
\end{equation*}
respectively. The $\mc T_4$-comodule structure now yields the module structure given by
\begin{align*}
    S_1 \act T_0 &\cong T_1 \cong \mathbb{C}\{v_1 \otimes m_0\} \, ,
    \\
    S_1 \act T_1 &\cong T_0 \cong \mathbb{C}\{v_1 \otimes m_1\} \, , 
\end{align*}
and
\begin{align*}
    P_0 \act T_0 & \cong T_0 \oplus T_1 
    \\[-.6em] &\cong 
    \mathbb C \{ v_0 \otimes m_0+ \frac{1}{2\sqrt{\xi}}v_1 \otimes m_0\} \oplus \mathbb C\{v_1 \otimes m_0\} \, ,
    \\
    P_0 \act T_1 &\cong T_0 \oplus T_1
    \\[-.6em] &\cong
    \mathbb C\{v_1 \otimes m_1 \} \oplus \mathbb C \{ v_0 \otimes m_1- \frac{1}{2\sqrt{\xi}}v_1 \otimes m_1\} \, ,
    \\
    P_1 \act T_0 &\cong T_0 \oplus T_1
    \\[-.6em] &\cong
    \mathbb C\{w_1 \otimes m_0\} \oplus \mathbb C \{w_0\otimes m_0 - \frac{1}{2\sqrt{\xi}}w_1\otimes m_0\} \, ,
    \\
    P_1 \act T_1 &\cong T_0 \oplus T_1
    \\[-.6em] &\cong
    \mathbb C \{w_0\otimes m_1 + \frac{1}{2\sqrt{\xi}}w_1\otimes m_1\} \oplus \mathbb C \{w_1 \otimes m_1\} \, .
\end{align*}

\bigskip \noindent
$\bul$ $\Mod(\mc A(2,\xi))$: Given the associative algebra structure \eqref{eq:AdXi}, it is found to be a rank 1 semisimple category whose unique simple object is the two-dimensional simple module
\begin{align*}
    T = \mathbb C\{m_0,m_1\} \q \text{w/} \;  
    \left\{
    \begin{array}{l}
        y \cdot m_0 = \sqrt \xi \, m_0   
        \\
        y \cdot m_1 = -\sqrt \xi \, m_1
        \\
        h \cdot m_0 = m_1
        \\
        h \cdot m_1 = m_0
    \end{array}
    \right. .
\end{align*}
The $\mc T_4$-comodule structure yields the module structure given by 
\begin{align*}
    S_0 \act T &\cong T \cong 
    \bC\{ w_1\otimes m_0 , w_1 \otimes m_1 \} \, ,
    \\
    S_1 \act T &\cong T
    \cong 
    \bC\{ v_1 \otimes m_1 , -v_1 \otimes m_0 \} \, ,
\end{align*}
and
\begin{align*}
    P_0 \act T \cong T \oplus T 
    \cong \; &\bC\{\frac{1}{2\sqrt{\xi}}v_1 \otimes m_0 + v_0 \otimes m_0, 
    \\[-.4em]
    & \q\, v_0 \otimes m_1-\frac{1}{2\sqrt{\xi}}v_1 \otimes m_1\} 
    \\  
    & \! \oplus \mathbb{C}\{v_1 \otimes m_1,- v_1 \otimes m_0\} \, ,
    \\ 
    P_1 \act T \cong T \oplus T 
    \cong \; &\bC\{w_0 \otimes m_1+\frac{1}{2\sqrt{\xi}}w_1 \otimes m_1, 
    \\[-.4em]
    & \q\, \frac{1}{2\sqrt{\xi}}w_1 \otimes m_0 - w_0 \otimes m_0\} 
    \\  
    & \! \oplus \mathbb{C}\{w_1 \otimes m_0,w_1 \otimes m_1\} \, .
\end{align*}

\subsection{$\F{\act}$-symbols}

\noindent
From here on, we use the notation $M_2 \in V \act M_1$ to mean $M_2$ appears as a summand in the decomposable object $V \act M_1$. 
From the various module structures  computed above, we can construct intertwining maps $\phi^{V \! M_1}_{M_2} \in \Hom_{\Mod(\mc A)}(V \act M_1,M_2)$ for each indecomposable object $M_2 \in V \act M_1$. Choosing bases $V = \mathbb C\{v_d\}_d$, $M_1 = \mathbb C\{m_{c_1}\}_{c_1}$ and $M_2 = \mathbb C\{m_{c_2}\}_{c_2}$, components of the linear map $\phi^{V \! M_1}_{M_2}$ are denoted by $\big(\phi^{V \! M_1}_{M_2} \big)_{d c_1}^{c_2} \in \mathbb C$. This allows us to define the following `action' tensors:
\begin{align}
    \actionTGen{\phi^{V \! M_1}_{M_2}}{}{}{}{}{} \!\! 
    &\equiv \!\!\! 
    \sum_{d, c_1,c_2} \!\!\! \actionTGen{\phi^{V \! M_1}_{M_2}}{d}{}{}{c_2\, }{\, c_1} \; m_{c_2} \otimes v_d^* \otimes m_{c_1}^*  
    \\
    &\equiv \!\!\! \sum_{d,c_1,c_2} \!\!\! \big(\phi^{V \! M_1}_{M_2} \big)_{dc_1}^{c_2} \; m_{c_2} \otimes v_d^* \otimes m_{c_1}^*   
    \, .
\end{align}
The module structures above also provide us with linear maps $\bar \phi^{V \! M_1}_{M_2} : M_2 \to V \act M_1$
satisfying orthogonality conditions
\begin{equation}
    \label{eq:orthoActionGen}
    \orthoActionGen = \delta_{M_2,M_3} \, \mathbb I_{M_2} \, .
\end{equation}
For instance, consider the algebra object $\mc A(1,\xi)$. By inspection, we find that the tensor \eqref{eq:actionT} introduced in the main text corresponds to the linear map
$\phi^\alpha \equiv \phi^{P_\alpha (T_0 \oplus T_1)}_{T_0 \oplus T_1} : P_\alpha \act (T_0 \oplus T_1) \to 2 \cdot (T_0 \oplus T_1)$ with $\alpha \in \{0,1\}$ under the identification
\begin{equation}
    \actionTGen{\;\;\, \phi^{P_\alpha \! T_{\gamma_1}}_{T_{\gamma_2}}}{}{}{}{}{} \!\!\!\! \equiv \;
    \actionT{\phi^\alpha}{\phantom{d}}{\gamma_2}{\gamma_1} \!\! ,
\end{equation}
for every $\gamma_1,\gamma_2 \in \{0,1\}$. We can now use this data, together with the isomorphism data in \eqref{eqn:H4projectivedecomp}, to explicitly compute the module associator of $\Mod(\mc A(1,\xi))$. 
The module associator boils to a collection of matrices of the form
\begin{equation}
\begin{split}
    \label{eq:Fmat}
    \F{\act}^{V_1 V_2 T_{\gamma_1}}_{T_{\gamma_2}} : 
    &\Hom_{\mc M}((V_1 \otimes V_2) \act T_{\gamma_1}, T_{\gamma_2})
    \\
    & \xrightarrow{\sim} \Hom_{\mc M}(V_1 \act (V_2 \act T_{\gamma_1}), T_{\gamma_2}) \, ,
\end{split}
\end{equation}
where $\mc M \equiv \Mod(\mc A(1,\xi))$, $V_1,V_2 \in \{S_0,S_1,P_0,P_1\}$ and $\gamma_1,\gamma_2 \in \{0,1\}$.
The fact that $\mc M$ is semisimple implies that these determine the full data of the natural transformation $\Hom_{\mc M}((V_1 \otimes V_2) \act T_{\gamma_1}, -) \xrightarrow{\sim} \Hom_{\mc M}(V_1 \act (V_2 \act T_{\gamma_1}), -)$, where an application of the \emph{Yoneda lemma} provides the inverse module associator isomorphism $\Fbar{\act}^{V_1 V_2 T_{\gamma_1}} : V_1 \act (V_2 \act T_{\gamma_1}) \xrightarrow{\sim} (V_1 \otimes V_2) \act T_{\gamma_1}$. Our choice of isomorphisms identifying $V_1 \otimes V_2$ and $V_2 \act T_{\gamma_1}$ with their respective decompositions into indecomposable objects provide the decompositions
\begin{equation}
\begin{split}
    &\Hom_{\mc M}((V_1 \otimes V_2) \act T_{\gamma_1}, T_{\gamma_2}) 
    \\
    & \q \cong
    \!\! \bigoplus_{V_3 \in V_1 \otimes V_2} \!\!
    \Hom_\mc M(V_3 \act T_{\gamma_1},T_{\gamma_2}) 
\end{split}
\end{equation}
and
\begin{equation}
\begin{split}
    &\Hom_{\mc M}(V_1 \act (V_2 \act T_{\gamma_1}), T_{\gamma_2}) 
    \\
    & \q \cong
    \!\! \bigoplus_{T_{\gamma_3} \in V_2 \! \act \! T_{\gamma_1}} \!\!
    \Hom_\mc M(V_1 \act T_{\gamma_3},T_{\gamma_2}) \, ,
\end{split}
\end{equation}
respectively. Together with our choice of intertwining maps $V_3 \act T_{\gamma_1} \to T_{\gamma_2}$ and $V_1 \act T_{\gamma_3} \to T_{\gamma_2}$, these decompositions provide bases for the hom-spaces appearing in eq.~\eqref{eq:Fmat}. It follows that we can view the linear map defined in eq.~\eqref{eq:Fmat} as a matrix indexed by $V_3 \in V_1 \otimes V_2$ and $T_{\gamma_3} \in V_2 \act T_{\gamma_1}$, the entries of which we refer to as $\F{\act}$-symbols. Carrying out the computations, we recover the $\F{\act}$-symbols computed in the main text via the identification
\begin{equation}
    \big(\F{\act}^{P_{\alpha_1} \! P_{\alpha_2} \! T_{\gamma_1}}_{T_{\gamma_2}}\big)^{T_{\gamma_3}}_{P_{\alpha_3}} \equiv \big(\F{\act}^{\alpha_1\alpha_2\gamma_1}_{\gamma_2}\big)^{\gamma_3}_{\alpha_3} \, ,
\end{equation}
for every $\gamma_1,\gamma_2,\gamma_3,\alpha_1,\alpha_2, \alpha_3 \in \{0,1\}$. Proceeding similarly, we can construct $F$-symbols associated with the monoidal structure of $\Mod(\mc T_4)$. One can then verify that the $\F{\act}$-symbols satisfy pentagon equations involving $F$-symbols, which are implied by pentagon equations satisfied by the monoidal associator.\footnote{We note two subtleties about $F$-symbols in the non-semisimple setting. Firstly, the decompositions that we use to define the $F$-symbols are in terms of indecomposable objects. For general finite tensor category, there may be infinitely many indecomposable objects, even though there are only finitely many simple and projective/injective objects. An alternative definition using only the projective objects was suggested in ref.~\cite{JubeirWang:2025reconftc}. Nonetheless, this subtlety does not apply to $\Mod(\mc T_4)$, as there are only finitely many indecomposable objects. Secondly, while $F$-symbols satisfying pentagon equations completely determine the monoidal associator in the semisimple setting, more conditions are required to ensure that the $F$-symbols lift to a monoidal associator in the non-semisimple setting.}

\subsection{Symmetric states\label{app:sym_States}}

\noindent
Given a finite semisimple indecomposable module  category $\mc M$ over a fusion category $\mc C$, vacua of the corresponding $\mc C$-symmetric gapped phase are in one-to-one correspondence with simple objects in $\mc M$ \cite{Thorngren:2019iar,Komargodski:2020mxz}. Moreover, representatives of the corresponding symmetric subspaces can be constructed from the data of these simple objects \cite{Inamura:2021szw,Bhardwaj:2024kvy}. Let us apply the same construction to exact indecomposable module categories over $\Mod(\mc T_4)$. 

Generally, let $\mc A = \mathbb C\{k_b\}_b$ be a right $\mc T_4$-simple left $\mc T_4$-comodule algebra so that $\Mod(\mc A)$ is an exact indecomposable left $\Mod(\mc T_4)$-module category. Let $\rho : \mc A \to \End(M)$ with $M = \mathbb C\{m_c\}_c$ be an indecomposable object in $\Mod(\mc A)$. Introducing the notation
\begin{equation}
    \rho_{c_1}^{c_2} := \sum_{b} \rho(k_b)_{c_1}^{c_2} \, k_b \, , 
\end{equation}
one defines a state in $\mc H_\Lambda := \bigotimes_{\msf i \in \Lambda} \msf{Forg}(\mc A)$ with open boundary conditions $m_{c_1}, m_{c_{|\Lambda|+1}} \in V$ as
\begin{equation}
    \sum_{c_2,\ldots,c_{|\Lambda|}} \rho_{c_{|\Lambda|}}^{c_{|\Lambda|+1}} \otimes \cdots \otimes \rho_{c_2}^{c_3} \otimes \rho_{c_1}^{c_2} \, .
\end{equation}
This state can be expressed as a tensor network of the form \eqref{eq:MPS} by defining the following tensor:
\begin{align}
    \nn
    \MPST{\rho}{}{}{}{}{} \!\!\! &\equiv \!
    \sum_{b,c_1,c_2} \! \MPST{\rho}{b}{}{}{\, c_1}{c_2 \,}  k_b \otimes m_{c_2}^* \otimes m_{c_1}
    \\
    \label{eq:MPST}
    &\equiv \!
    \sum_{b,c_1,c_2} \! \rho(k_b)_{c_1}^{c_2} \;  k_b \otimes m_{c_2}^* \otimes m_{c_1} \, .
\end{align}
Repeating this procedure for every indecomposable object in $\Mod(\mc A)$ yields states that span a subspace of $\mc H_\Lambda$. We claim that these states are symmetric in the sense that the subspace they span is invariant under the action of the symmetry $\Mod(\mc T_4)$, as defined in sec.~\ref{app:chain}, for every finite subset $\Lambda$ of $\mathbb Z$. Indeed, let $\rho_{M_1} : \mc A \to \End(M_1)$ be an indecomposable object in $\Mod(\mc A)$ and $\varrho : \mc T_4 \to \End(V)$ be an indecomposable object in $\Mod(\mc T_4)$. By definition of the $\Mod(\mc T_4)$-module structure of $\Mod(\mc A)$, the intertwining map $\phi^{V\! M_1}_{M_2} : V \act M_1 \to M_2$, for any $M_2 \in V \act M_1$ satisfies
\begin{equation}
\begin{split}
    &\sum_{b_2,c_2,d_2} \big(\phi_{M_2}^{V \! M_1}\big)_{d_2 c_2}^{c_3}  \, \varrho(\lambda_{b_1}^{b_2})_{d_1}^{d_2} \, \rho_{M_1}(k_{b_2})_{c_1}^{c_2} 
    \\
    & \q = 
    \sum_{c_2,d_2} \big(\phi_{M_2}^{V \! M_1}\big)_{d_2 c_2}^{c_3}  \, (\varrho \act \rho_{M_1})\big(\lambda(k_{b_1})\big)_{d_1c_1}^{d_2c_2}
    \\
    & \q = \sum_{c_4} \rho_{M_2}(k_{b_1})^{c_3}_{c_4} \, \big(\phi^{V \! M_1}_{M_2} \big)_{d_1c_1}^{c_4} \, ,
\end{split}
\end{equation}
from which the following tensor network identity follows:
\begin{equation}
    \raisebox{12.5pt}{\actionMPO{1}{\rho_{M_1}}{\omega_V}{}{}{} } = \!\!
    \sum_{M_2 \in V \! \act \! M_1} \!\!\!\!\!
    \raisebox{4.8pt}{\actionMPO{2}{\rho_{M_2}}{}{\;\;\;\;\, \phi^{V \! M_1}_{M_{2_{\phantom L}}} }{\bar \phi^{V \! M_1}_{M_{2_{\phantom L}}} \;\;\;}{} } \hspace{-10pt} .
\end{equation}
Together with orthogonality condition \eqref{eq:orthoActionGen}, this guarantees that the subspace spanned by states labelled by every indecomposable object in $\Mod(\mc A)$ is indeed invariant under $\Mod(\mc T_4)$. Let us now examine explicit examples: 

\bigskip\noindent
$\bul$ $\Mod(\mc A(1,0))$: In app.~\ref{app:moduleCat}, we computed two indecomposable modules over $\mc A(1,0)$, namely the simple module $S$ and its projective cover $P$. On the one hand, it readily follows from the definition that the state in $\mc H_\Lambda$ associated with $S$ is $|0\ra^{\otimes |\Lambda|}$ under the identification $1 \equiv k_0 \equiv |0\ra$. On the other hand, choosing the projective object $P$, we recover the W state $|\text{W}\ra_\Lambda$ for the choice of open boundary conditions depicted in eq.~\eqref{eq:MPSW}, under the identifications $1 \equiv |0\ra$ and $y \equiv |1\ra$. Note that choosing different boundary conditions results either in the product state $|0\ra^{\otimes |\Lambda|}$ or the zero state. The fact that acting on the product state with the symmetry operator $(\widehat \omega_0)_0^1$ defined in the main text results in the W state is now explained by $P_0 \act S \cong P$.

The most interesting feature of this example is the non-identity map $P \to P$ that factors through $S$. This map indicates the existence of topological local operators locally turning the W state into the product state. Indeed, consider the maps $\pi : P \to S$, $m_0 \mapsto m_1$, $m_1 \mapsto 0$ and $\iota : 
S \to P$, $m_1 \mapsto m_1$. These can be used to locally modify the W state as follows:
\begin{equation}
\begin{split}
    &\MPSW{1} 
    \\ & \q = \MPSW{2} \, ,
\end{split}
\end{equation}
where the equality follows from the  topological invariance of the local operators, at which point it becomes indistinguishable from the product state. 
Therefore, even though the indecomposable objects provide two distinct ground states breaking the $\Mod(\mc T_4)$ symmetry down to $\mathbb Z / 2 \mathbb Z$, they should correspond to the same vacuum in the infrared limit. Finally, note that one can replace the insertion of the matrices $\pi$ and $\iota$ above, by a single matrix labelled by $\iota \circ \pi = \sigma^-$. The insertion of such a matrix within the tensor network amounts to acting on the W state with the physical operator $\sigma^+_\msf i : \mc H_\Lambda \to \mc H_\Lambda$.

\bigskip\noindent
$\bul$ $\Mod(\mc A(1,\xi))$: In app.~\ref{app:moduleCat}, we computed two indecomposable modules over $\mc A(1,\xi)$, namely the simple modules $T_0$ and $T_1$. It readily follows from the definitions that applying our construction to these two modules yields the product states $|+\xi \ra^{\otimes |\Lambda|}$ and $|-\xi \ra^{\otimes |\Lambda|}$ considered in the main text, respectively. There, we defined in eq.~\eqref{eq:MPSXM} a tensor of the form \eqref{eq:MPST} associated with $M= T_0 \oplus T_1$ so as to be able to compute the action of $\Mod(\mc T_4)$ on an arbitrary state in the invariant subspace. In particular, the isomorphisms $S_1 \act T_{0/1} \cong T_{1/0}$ and $P_0 \act T_{0/1} \cong T_0 \oplus T_1$ explain the action of the symmetry operators on the ground states. In light of this identification, we can now confirm that the $\F{\act}$-symbols computed in the main text do specify the module associator of the $\Mod(\mc T_4)$-module category $\Mod(\mc A(1,\xi))$. In this context, the fact that ground states of the Hamiltonian $\widetilde{\mathbb H}(\xi)_\Lambda$ were found to transform inequivalently under $\Mod(\mc T_4)$ follows from the fact that the corresponding $\Mod(\mc T_4)$-module categories $\Mod(\mc A(1,\xi))$ are inequivalent.

\bigskip\noindent
$\bul$ $\Mod(\mc A(2,\xi))$: In contrast to the two previous cases, we found a single simple module over $\mc A(2,\xi)$, for any $\xi \in \mathbb C^\times$. Whenever $\xi = 1$, we recognise the corresponding tensor network state as the so-called \emph{cluster state} \cite{PhysRevLett.86.910}. Although the state is here found to be $\Mod(\mc T_4)$-symmetric, it is typically defined as the unique symmetric ground state of a $\mathbb Z/2 \mathbb Z \times \mathbb Z / 2 \mathbb Z$-symmetric Hamiltonian. In regard to this $\mathbb Z / 2 \mathbb Z \times \mathbb Z / 2 \mathbb Z$ symmetry, we commented in app.~\ref{app:algObj} that as an algebra $\mc A(2,\xi)$ is isomorphic to the group algebra $\mathbb C[\mathbb Z/2 \mathbb Z \times \mathbb Z / 2 \mathbb Z]$ twisted by the 2-cocycle $\beta_\xi$---here interpreted as a group cocycle. 
But, for $\xi \neq \xi'$, the group 2-cocycles $\beta_\xi$ and $\beta_{\xi'}$ fall within the same cohomology class. 
Computing the local action of the symmetry $\mathbb Z/2 \mathbb Z \times \mathbb Z / 2 \mathbb Z$ on the corresponding states would thus reveal that they all transform equivalently, up to gauge transformations of the action tensors. However, repeating the analysis of the main text for this example would reveal that the states transform inequivalently with respect to the $\Mod(\mc T_4)$-symmetry. Once we have established that the invariant $\F{\act}$-symbols specify the module associator of the $\Mod(\mc T_4)$-module category $\Mod(\mc A(2,\xi))$, this follows from the fact that Hopf 2-cocycles $\beta_\xi$ and $\beta_{\xi'}$ fall within distinct equivalence classes, for any $\xi \neq \xi'$.  

\bigskip \noindent
Below, we discuss how to construct parent Hamiltonians for the aforementioned symmetric states. 

\subsection{Parent Hamiltonians\label{app:parent}}

\noindent
Given a symmetry (unitary) fusion category $\mc C$ admitting a fibre functor and  a choice of indecomposable finite semisimple $\mc C$-module category $\mc M$ encoding a gapped $\mc C$-symmetric phase, a commuting projector Hamiltonian representing the gapped phase can be defined from the data of a $\Delta$-separable symmetric Frobenius algebra object $\mc A$ in $\mc C$ such that $\Mod_\mc C(\mc A) \simeq \mc M$ \cite{Inamura:2021szw,Bhardwaj:2024kvy}.
Crucially, given an indecomposable, semisimple, unital, associative algebra in $\mc C$, it is always possible to endow it with a $\Delta$-separable symmetric Frobenius structure \cite{Fuchs:2002cm,fuchs2009,kong2019}. However, this procedure generally fails in the case of a finite tensor category. 
In our case, this failure can be traced back to the fact that the algebra object $\mc A$ may not possess a non-zero map into the monoidal unit, which already obstructs the existence of a counit. 
It implies that we may not always be able to construct a gapped symmetric commuting projector self-adjoint parent Hamiltonian, but we may still 
be able to construct---at least in some cases---a gapped symmetric parent Hamiltonian with a real spectrum.

Let us specialise immediately to the case of $\mc A(1,\xi)$ with $\xi \in {\rm U}(1)$. We are looking for a parent Hamiltonian for the ground states labelled by the indecomposable objects in $\Mod(\mc A(1,\xi))$.
Following the discussion at the end of app.~\ref{app:chain}, we can turn any morphism $\mc A(1,\xi)^{\otimes 2} \to \mc A(1,\xi)^{\otimes 2}$ in $\Comod(\mc T_4)$ into a local symmetric operator $\mathbb C^2 \otimes \mathbb C^2 \to \mathbb C^2 \otimes \mathbb C^2$ acting on neighbouring sites of the microscopic Hilbert space. We choose to write this morphism as a composition 
\begin{equation}
    \Delta \circ \mu :\mc A(1,\xi)^{\otimes 2} \to \mc A(1,\xi) \to \mc A(1,\xi)^{\otimes 2} \, ,
\end{equation}
where $\mu$ is the multiplication in $\mc A(1,\xi)$ and $\Delta : \mc A(1,\xi) \to \mc A(1,\xi)^{\otimes 2}$ is a morphism that remains to be determined. By definition, a morphism $\Delta : \mc A(1,\xi) \to \mc A(1,\xi)^{\otimes 2}$ in $\Comod(\mc T_4)$ consists of a morphism in $\Vect$ satisfying 
\begin{equation}
    \label{eq:AAComod}
    (\text{id}_{\mc T_4} \otimes \Delta) \circ \lambda = \lambda_{\mc A(1,\xi)^{\otimes 2}} \circ \Delta \, ,
\end{equation}
where $\lambda_{\mc A(1,\xi)^{\otimes 2}} : \mc A(1,\xi)^{\otimes 2} \to \mc \mc \mc T_4 \otimes \mc A(1,\xi)^{\otimes 2}$, $a \otimes b \mapsto (a_{(-1)} \cdot b_{(-1)}) \otimes a_{(0)} \otimes b_{(0)}$, for every $a,b \in \mc A(1,\xi)$, endows $\mc A(1,\xi)^{\otimes 2}$ with a $\mc T_4$-comodule structure. We find a two-dimensional space of morphisms $\Delta$ satisfying eq.~\eqref{eq:AAComod} defined by
\begin{equation}
\begin{split}
    \Delta(1) &:= \zeta_1 (1 \otimes 1) 
    \\
    \Delta(y) &:=  \zeta_2 (1 \otimes y) + (\zeta_1 - \zeta_2) (y \otimes 1) \, ,
\end{split}
\end{equation}
for every $\zeta_1, \zeta_2 \in \mathbb C$. Now, in order to construct a \emph{frustration free} parent Hamiltonian for the states defined in app.~\ref{app:sym_States} associated with $T_0,T_1 \in \Mod(\mc A(1,\xi))$, it is sufficient to require $\mu \circ \Delta = \text{id}_{\mc A(1,\xi)}$, which in turn forces $\zeta_1 = 1$, while we are still free to choose $\zeta_2$. Requiring $\Delta$ to be \emph{coassociative} further restricts $\zeta_2$ to $\{0,1\}$. Without loss of generality, we choose $\zeta_2=0$. Bringing everything together, let us consider the morphism $\mc A(1,\xi)^{\otimes 2} \to \mc A(1,\xi)^{\otimes 2}$
\begin{equation}
\begin{split}
    \Delta \circ \mu : 
    1 \otimes 1 &\mapsto 1 \otimes 1
    \\
    1 \otimes y &\mapsto y \otimes 1 
    \\
    y \otimes 1 &\mapsto y \otimes 1
    \\
    y \otimes y &\mapsto \xi \cdot 1 \otimes 1
\end{split}\, .
\end{equation}
We finally define the local operator $\widetilde{\mathbb h}(\xi)_{\msf i, \msf i+1}$ as the embedding of the \emph{transpose} of $\msf{Forg}(\Delta \circ \mu)$ into the microscopic Hilbert space $\mc H_\Lambda$.\footnote{As in eq.~\eqref{eq:MPOT}, the transpose is merely for the need of our exposition in the main text.} Under the identifications $1 \equiv |0\ra$ and $y \equiv |1\ra$, one recovers local operators \eqref{eq:localOpXM}, as expected. Following the same steps for $\mc A(1,0)$, one recovers $\widetilde{\mathbb h}(0)_{\msf i, \msf i+1}$. The same strategy also produces a parent Hamiltonian for the ground state associated with the unique simple object in $\Mod(\mc A(2,\xi))$, for every $\xi \in \mathbb C^\times$, however the spectrum of this Hamiltonian does not appear to be real. 

The construction provided above can also be used to justify local operators \eqref{eq:localOpXX}. We explained in app.~\ref{app:sym_States} how ground states $| +\xi \ra^{\otimes |\Lambda|}$ and $|-\xi\ra^{\otimes |\Lambda|}$ can be constructed from the simple modules $T_0$ and $T_1$ over $\mc A(1,\xi)$, respectively. Consider instead the twisted group algebra $\mathbb C[\mathbb Z / 2 \mathbb Z]^{\beta_\xi} = \mathbb C\la h \, | \, h^2 = \xi \cdot 1 \ra$, where $\beta_\xi$ is treated here as a 1-coboundary of $\mathbb Z / 2\mathbb Z$. Since $\mathbb C[\mathbb Z / 2 \mathbb Z]^{\beta_\xi}$ is isomorphic to $\mc A(1,\xi)$ as an algebra, they share the same simple modules. However, they do not possess the same $\mc T_4$-comodule structure, so that the map $\Delta$ is now defined by $\Delta(1) = \frac{1}{2\xi}(\xi \cdot 1 \otimes 1 + h \otimes h)$ and $\Delta(h)=\frac{1}{2}(1 \otimes h + h \otimes 1)$. Proceeding as before, one recovers local operators \eqref{eq:localOpXX} under the identification $1 \equiv |0\ra$ and $h \equiv |1\ra$.

\newpage

\newpage
\bibliography{refs}

\end{document}